\documentclass[journal]{IEEEtran}

\usepackage{amssymb}
\usepackage{cite}
\usepackage{amsmath}
\usepackage{xcolor}
\usepackage{multirow}
\usepackage{float}
\usepackage{colortbl}
\usepackage{booktabs}
\usepackage{array}
\usepackage{mathtools, cuted}
\usepackage{makecell}
\usepackage{amsmath}%
\usepackage{enumitem}
\usepackage{dblfloatfix}
\usepackage{graphicx}

\definecolor{mycolor}{rgb}{.949,.949,.949}

\usepackage{hyperref} 
\hypersetup{
    colorlinks=true,       
    linkcolor=blue,        
    citecolor=blue,        
    urlcolor=blue,         
    pdfborder={0 0 0}      
}

\usepackage{algorithm}
\usepackage{algorithmic}

\usepackage{tikz}

\newcommand{\mycomment}[1]{}

\begin{document}
\bstctlcite{IEEEexample:BSTcontrol}

\title{Revisiting the Effect of Grid-Following Converter on Frequency Dynamics - Part II: Spatial Variation}
\author{Jiahao~Liu,~\IEEEmembership{Student Member,~IEEE}, Cheng~Wang,~\IEEEmembership{Senior Member,~IEEE}, Tianshu~Bi,~\IEEEmembership{Fellow,~IEEE}
}

\maketitle

\begin{abstract}
    Besides the center of inertia (COI) frequency dynamics addressed in Part I, the spatial frequency variation in power systems with grid-following (GFL) converters is also crucial. Part II revisits the effect of GFLs on frequency spatial variation. Leveraging the interfacing state variables and equivalent frequency defined in Part I, an extended frequency divider (FD) formula is proposed. The linearized mapping relationship between network node frequency and synchronous generator (SG) rotor frequency, as well as GFL equivalent frequency, is modeled. The superposition contribution from GFLs is determined by the electrical distance between the generator and the frequency observation node, as well as the system power flow conditions. Additionally, the frequency mapping for branch currents, which is overlooked in previous research, is addressed. Simulation results validate the accuracy of the proposed extended FD formula. They quantitatively demonstrate that the superposition contribution of GFLs to node frequency is relatively weak and that the superposition coefficient is time-varying. The branch frequency superposition reveals a complex and distinctly different pattern.
\end{abstract}

\section{Introduction\label{Sec:Introd}}

\IEEEPARstart{T}HE complex differential-algebraic equations (DAEs) of power systems allows system frequency to exhibit spatial differences during disturbances \cite{BTan22}. The frequency measured at different locations in the system varies, even though these variations generally follow the trend of the center of inertia (COI) frequency \cite{XChen24}. Understanding frequency spatial variation is critical because system operation and control rely on local frequency. For instance, phasor measurement units (PMUs) and frequency-triggered protection devices are installed and operate locally \cite{YLiu16, MWAltaf22, MSun21}.

In traditional power systems, the mechanisms governing frequency distribution are well understood. According to the synchronizing torque and the swing equation of synchronous generators (SGs), the frequency at network is determined by the rotor frequencies of all generators through a superposition principle \cite{FMilano18Review}. The contribution of each SG to the local frequency superposition is influenced by the electrical distance between the measurement location and the generator. This relationship is quantified by the well-known Frequency Divider (FD) formula, which is derived from the system's DAEs \cite{FMilano17}.

The inclusion of inverter-based resources (IBRs), especially grid-following (GFL) converters, adds complexity to the understanding of frequency spatial mechanisms. GFLs are synchronized via phase-locked loop (PLL) and operate with their own complex control logic \cite{XHe21}. The only widely accepted understanding is that the PLL passively tracks the terminal frequency and has a limited impact on network frequency compared to SGs \cite{FMilano17}. Given this ambiguity, it is essential to thoroughly investigate the effect of GFLs on frequency spatial variation.

The existing perspectives on this topic can be categorized into three primary approaches.

The first is the \textbf{over-simplified approach}. In \cite{FMilano17}, it is claimed that GFLs impose no impact on network frequency since their dynamics are entirely ignored. As a result, the traditional FD formula can still be applied with high accuracy. Following this perspective, numerous works on power system frequency stability research with GFL penetration have emerged. Notable examples include system modeling that accounts for frequency spatial differences \cite{FMilano19}, frequency control parameter placement based on nodal frequency \cite{JZhang24}, and real-time frequency control strategy \cite{AHussain21}. This approach cannot be applied to systems with higher IBR penetration because GFLs inevitably exhibit their own dynamics.

The second is the \textbf{GFL-amended approach}. In this approach, the impact of GFLs on network frequency is considered, acknowledging that GFLs, especially through their PLL, exhibit their own dynamics \cite{QMa23, QMa24, BTan2022FDF}. Quantification is achieved by first modeling the impact of GFLs on their terminal frequency. The mapping is then incorporated into the original FD formula by treating GFLs as SGs. For instance, \cite{QMa23} models the mapping relationship between GFL state variables and terminal voltage, deriving its contribution to terminal frequency through time differentiation. Similarly, \cite{BTan2022FDF} uses impedance modeling to establish the relationship between the virtual frequency of GFLs and terminal frequency. These methods show improvements compared to the original FD formula. However, existing methods primarily focus on the impact of GFLs at their terminals, rather than addressing their influence on the entire system dynamics directly.

The third is the recently proposed \textbf{complex frequency approach}. This approach offers a valuable insight that any device capable of altering its active and reactive power can influence the nodal frequency, which is represented as a complex variable \cite{FMilano22}. The above relationship is quantitatively mapped and transformed into a mapping between the current injected into each node and the nodal complex frequency. GFLs can be analyzed using this method, as demonstrated for GFLs with PLLs \cite{DMoutevelis24} and virtual frequency regulation \cite{FMilano20pt1, FMilano20pt2}. However, as highlighted in \cite{FMilano22}, this approach requires generators to be modeled as current sources, with their injected current also treated as a state variable. This requirement limits its application in comparing the mechanisms by which SGs and GFLs impact frequency dynamics.

Considering the ambiguity in GFL research, it is essential to revisit how SGs influence frequency spatial characteristics. Network frequency, as an algebraic variable, is determined by the system's state variables. When considering only the second-order model of SGs, the key state variable is the electromotive force (EMF) angle of the SG, along with its time derivative, the rotor frequency. Since rotor frequencies differ across the system, network frequency exhibits spatial differences. This relationship is quantified using the FD formula in \cite{FMilano17}, which is derived from the network equations of system DAEs. The initial version of the ``frequency divider'' proposed in \cite{JNutaro12} considers state variables to include the EMF amplitude. As a result, the time derivative of the EMF amplitude, alongside the angle part, which represents rotor frequency, is recognized as a contributing factor to network frequency. More advanced quantifications based on network equations are presented in \cite{JNutaro12}. From this, it is evident that the core of frequency spatial variation analysis lies in identifying the system's state variables and linking them to network frequency.

In Part II of this series, the impact of GFLs on spatial frequency variation is revisited, leveraging the internal state variables of GFLs interfacing with power systems as defined in Part I. Compared to existing methods, this work makes three key contributions:

(1) The extended frequency divider (FD) formula is proposed, establishing the mapping relationships between network node frequency and the state variables of SGs and GFLs. Superposition coefficients from GFL equivalent frequencies, derived from the angle and amplitude parts of its state variables, are explicitly provided. This formulation avoids any approximations, ensuring an accurate representation of node frequency.

(2) Leveraging the modeled linearized relationship, the effect of GFL equivalent frequency on network node frequency is analyzed quantitatively. This impact is influenced by the electrical distance between the GFL and the observation node, which is determined by network topology and parameters. Additionally, the effect is dependent on power flow conditions. Overall, the contribution from GFLs is found to be significantly weaker compared to SGs.

(3) An additional problem, overlooked in existing literature, is identified. The spatial patterns of frequencies measured at network node voltages and branch currents, both of which are outputs of PMUs according to IEEE standard \cite{IEEEC37}, are found to be fundamentally different. Based on the derived extended FD formula for branch frequency, it is unexpectedly revealed that branch frequency is not bounded by either SG rotor frequency or GFL equivalent frequency.

\section{Problem Formulation\label{Sec:ProbForm}}

\subsection{Frequency Spatial Variation and Its Quantification}

In traditional power systems, the network node frequency is determined by the frequencies of all SGs in the system. A well-known approach for linearly quantifying the superposition relationship between SG frequencies and node frequency is the FD formula proposed in \cite{FMilano17}, which is expressed as:
\begin{equation}\label{eq:TdFDF}
    \Delta \boldsymbol{\omega}^{N} = \boldsymbol{D} \Delta \boldsymbol{\omega}^{G},
\end{equation}
where $\boldsymbol{\omega}^{N} \in \mathbb{R}_{N^{N}\times 1}$ and $\boldsymbol{\omega}^{G} \in \mathbb{R}_{N^{G}\times 1}$ represent the vectors of network node frequencies and SG rotor frequencies, respectively, with elements $\omega^{N}$ and $\omega^{G}$; $N^{N}$ and $N^{G}$ denote the total number of network nodes and SGs in the system; the superposition matrix $\boldsymbol{D}$ is determined by the network topology and the system reactance parameters, as detailed in \cite{FMilano17}.

\subsection{Tasks for GFL Study\label{SubSec:Tasks}}

Before clarifying the research tasks for GFLs, it is important to delve deeper into frequency spatial in SG-dominated system. In fact, (\ref{eq:TdFDF}) is insufficient even in SG-dominated system. The node frequency is determined by the state variables on the SG side that operate within the system frequency time scale. For instance, the angle and amplitude of the EMF of SGs, denoted as $E^{G}$ and $\delta^{G}$, respectively, also contribute to the node frequency when SG models beyond the second order are considered \cite{JNutaro12}.

Thus, in systems with GFLs, the state variables of GFLs within the system frequency time scale also contribute to the network frequency. Considering the GFL module diagram provided in Fig. 3 and the assumptions made in Section III of Part I of these series papers, the tasks for quantifying the effect of GFLs on frequency spatial variation include the following:
\begin{itemize}
    \item \textbf{Deriving an extended FD formula}: First, the equivalent frequency on the GFL side within the system frequency dynamics time scale needs to be clarified. Then, an extended FD formula, in a linearized form similar to (\ref{eq:TdFDF}), should be derived based on the system network topology and parameters.
    \item \textbf{Analyzing the effect}: Once the extended FD formula is established, the mechanism by which GFLs influence frequency spatial variation can be analyzed quantitatively.
\end{itemize}

It should also be noted that virtual frequency regulation and voltage droop control functions, as shown in papers like \cite{MZhang18}, are not depicted in Fig. 3 of Part I. However, the proposed method remains applicable to these scenarios because such functions do not alter the GFL state variables interfacing with the system.

\section{Extended FD Formula Including GFLs\label{Sec:ProFDF}}

The definition of the frequency source on the GFL side is clarified in Section \ref{SubSec:VirtFreq}. The extended FD formula for node voltage frequency is derived in Section \ref{SubSec:ExtFDFV}, and then extended to branch current frequency in Section \ref{SubSec:ExtFDFI}.

\subsection{Equivalent Frequency Source of GFL\label{SubSec:VirtFreq}}

The source of frequency on the SG side is mainly the SG rotor frequency, $\omega^{G}$, which is derived from the time derivative of SG state variables operating within the system frequency dynamics time scale, specifically the EMF, $E^{G} \angle \delta^{G}$, in the second-order modeling of SGs. Before defining the frequency source of GFLs, it is essential to identify the relevant state variables. As defined in Part I of this paper series, the state variable interfacing with the system network is given as:
\begin{equation}\label{eq:DefSSF}
    \text{interface state variable: } I^{F} \angle \theta^{F,I},
\end{equation}
where $I^{F}$ and $\theta^{F,I}$ represent the amplitude and angle of the GFL output current, respectively.

Based on (\ref{eq:DefSSF}), the frequency source is derived from the time derivative of the angle $\theta^{F,I}$, referred to as the equivalent frequency of the GFL, and is expressed as:
\begin{equation}\label{eq:SSDynW_w}
    \Delta \omega^{F}= \frac{\mathrm{d} \theta^{F,I}}{\mathrm{d} t} = c^{Pi} \Delta \omega^{F,Id} + \Delta \omega^{F,Pll},
\end{equation}
where the total equivalent frequency consists of two components, $\omega^{F,Id}$ and $\omega^{F,Pll}$, as defined in Part I; the value of the linearization coefficient $c^{Pi}$ is provided in Part I.

Symmetrically, another potential equivalent frequency source is the derivative of the amplitude $I^{F}$, represented as $\mathrm{d} I^{F} / \mathrm{d} t$. As demonstrated in the following sections, the contribution of this term is very small.

\subsection{Extended FD Formula\label{SubSec:ExtFDFV}}

The derivation begins with the system network equation, which connects all state variables on the SG side, the GFL side, and the node voltage phasor. Let the state variables $E^{G} \angle \delta^{G}$, $I^{F} \angle \theta^{F,I}$, and the network node voltage $U^{N} \angle \theta^{N}$ be represented in phasor form as $\dot{E}^{G}$, $I^{F}$, and $U^{N}$, respectively. The network equation is given here:
\begin{equation}\label{eq:IYU}
    \left[\begin{matrix}
        \boldsymbol{\dot{I}}^{G} \\
        \boldsymbol{\dot{I}}^{F} \\
        \boldsymbol{\dot{I}}^{N}
    \end{matrix}\right] = 
    \left[\begin{matrix}
        \boldsymbol{\dot{Y}}^{GG} & \boldsymbol{\dot{Y}}^{GF} & \boldsymbol{\dot{Y}}^{GN} \\
        \boldsymbol{\dot{Y}}^{FG} & \boldsymbol{\dot{Y}}^{FF} & \boldsymbol{\dot{Y}}^{FN} \\
        \boldsymbol{\dot{Y}}^{NG} & \boldsymbol{\dot{Y}}^{NF} & \boldsymbol{\dot{Y}}^{NN}
    \end{matrix}\right] 
    \left[\begin{matrix}
        \boldsymbol{\dot{E}}^{G} \\
        \boldsymbol{\dot{U}}^{F} \\
        \boldsymbol{\dot{U}}^{N}
    \end{matrix}\right],
\end{equation}
where the variables are rewritten in vector form as $\boldsymbol{\dot{E}}^{G} \in \mathbb{C}_{N^{G} \times 1}$, $\boldsymbol{\dot{I}}^{F} \in \mathbb{C}_{N^{F} \times 1}$, and $\boldsymbol{\dot{U}}^{N} \in \mathbb{C}_{N^{N} \times 1}$, where $N^{F}$ is the total number of GFLs in the system; while $\dot{\boldsymbol{I}}^{G}$, $\dot{\boldsymbol{U}}^{F}$, and $\dot{\boldsymbol{I}}^{N}$ represent the SG current, GFL terminal voltage, and current injection at network node in phasor form, respectively; the admittance matrix is denoted as $\boldsymbol{\dot{Y}}$.

To bring all state variables to the same side of the equation, (\ref{eq:IYU}) is rearranged as follows:
\begin{equation}\label{eq:IYU_re}
    \left[\begin{matrix}
        \boldsymbol{\dot{I}}^{G} \\
        \boldsymbol{\dot{U}}^{F} \\
        \boldsymbol{\dot{I}}^{N}
    \end{matrix}\right] = 
    \left[\begin{matrix}
        \boldsymbol{\dot{Y}}^{eq,GG} & \boldsymbol{\dot{T}}^{eq,GF} & \boldsymbol{\dot{Y}}^{eq,GN} \\
        \boldsymbol{\dot{T}}^{eq,FG} & \boldsymbol{\dot{Z}}^{eq,FF} & \boldsymbol{\dot{T}}^{eq,FN} \\
        \boldsymbol{\dot{Y}}^{eq,NG} & \boldsymbol{\dot{T}}^{eq,NF} & \boldsymbol{\dot{Y}}^{eq,NN}
    \end{matrix}\right] 
    \left[\begin{matrix}
        \boldsymbol{\dot{E}}^{G} \\
        \boldsymbol{\dot{I}}^{F} \\
        \boldsymbol{\dot{U}}^{N}
    \end{matrix}\right],
\end{equation}
where the symbols in the connection matrix are aligned with their physical meanings: $\dot{Y}^{eq}$ and $\dot{Z}^{eq}$ represent the equivalent admittance and impedance, respectively, while $\dot{T}^{eq}$ is defined as the transfer parameter between currents or between voltages. The detailed expressions for these coefficients in the first two rows are provided in the Appendix \cite{AddDoc}, and the last row is given here:
\begin{subequations}\label{eq:CoefYTZ}
\begin{equation}\label{eq:CoefYTZ_NG}
    \boldsymbol{\dot{Y}}^{eq,NG} = \boldsymbol{\dot{Y}}^{NG} - \boldsymbol{\dot{Y}}^{FG} \boldsymbol{\dot{Y}}^{NF} {\boldsymbol{\dot{Y}}^{FF}}^{-1},
\end{equation}
\begin{equation}\label{eq:CoefYTZ_NF}
    \boldsymbol{\dot{T}}^{eq,NF} = \boldsymbol{\dot{Y}}^{NF} {\boldsymbol{\dot{Y}}^{FF}}^{-1},
\end{equation}
\begin{equation}\label{eq:CoefYTZ_NN}
    \boldsymbol{\dot{Y}}^{eq,NN} = \boldsymbol{\dot{Y}}^{NN} - \boldsymbol{\dot{Y}}^{FN} \boldsymbol{\dot{Y}}^{NF} {\boldsymbol{\dot{Y}}^{FF}}^{-1}.
\end{equation}
\end{subequations}

Considering that the load connected at network nodes is represented by its admittance or impedance, the node current $\boldsymbol{\dot{I}}^{N}$ can be eliminated in (\ref{eq:IYU_re}). Using admittance as an example, (\ref{eq:IYU_re}) can be rewritten as follows: \footnote{In \cite{FMilano17}, the current $\boldsymbol{\dot{I}}^{N}$ is directly neglected. However, this assumption does not hold here because the order of magnitude of $\boldsymbol{\dot{U}}^{F}$ is significantly smaller than that of $\boldsymbol{\dot{I}}^{N}$}:
\begin{equation}\label{eq:IYU_lo}
    \left[\begin{matrix}
        \boldsymbol{\dot{I}}^{G} \\
        \boldsymbol{\dot{U}}^{F} \\
        \boldsymbol{0}
    \end{matrix}\right] = 
    \left[\begin{matrix}
        \boldsymbol{\dot{Y}}^{eq,GG} & \boldsymbol{\dot{T}}^{eq,GF} & \boldsymbol{\dot{Y}}^{eq,GN} \\
        \boldsymbol{\dot{T}}^{eq,FG} & \boldsymbol{\dot{Z}}^{eq,FF} & \boldsymbol{\dot{T}}^{eq,FN} \\
        \boldsymbol{\dot{Y}}^{eq,NG} & \boldsymbol{\dot{T}}^{eq,NF} & \boldsymbol{\dot{Y}}^{eq\prime,NN}
    \end{matrix}\right] 
    \left[\begin{matrix}
        \boldsymbol{\dot{E}}^{G} \\
        \boldsymbol{\dot{I}}^{F} \\
        \boldsymbol{\dot{U}}^{N}
    \end{matrix}\right],
\end{equation}
where $\boldsymbol{\dot{Y}}^{eq\prime,NN} = \boldsymbol{\dot{Y}}^{eq,NN} - \Lambda \left( \dot{Y}^{N} \right) $, and $\dot{Y}^{N}$ is the load equivalent admittance. Note that for nodes without a connected load, $\dot{Y}_{L}=0$.

By extracting the last column of (\ref{eq:IYU_lo}), the following node voltage superposition correlation can be obtained:
\begin{subequations}\label{eq:USS}
\begin{equation}\label{eq:USS_U}
    \boldsymbol{\dot{U}}^{N} = \left[ \begin{matrix} 
        \boldsymbol{\dot{D}}^{G} & \boldsymbol{\dot{D}}^{F}
    \end{matrix} \right]
    \left[\begin{matrix}
        \boldsymbol{\dot{E}}^{G} \\
        \boldsymbol{\dot{I}}^{F}
    \end{matrix}\right],
\end{equation}
\begin{equation}\label{eq:USS_DG}
    \boldsymbol{\dot{D}}^{G} = 
    - {\boldsymbol{\dot{Y}}^{eq\prime,NN}}^{-1} 
    \boldsymbol{\dot{Y}}^{eq,NG},
\end{equation}
\begin{equation}\label{eq:USS_DF}
    \boldsymbol{\dot{D}}^{F} = 
    - {\boldsymbol{\dot{Y}}^{eq\prime,NN}}^{-1} 
    \boldsymbol{\dot{T}}^{eq,NF},
\end{equation}
\end{subequations}
where $\boldsymbol{\dot{D}}^{G} \in \mathbb{R}_{N^{N} \times N^{G}}$ and $\boldsymbol{\dot{D}}^{F} \in \mathbb{R}_{N^{N} \times N^{F}}$ are the voltage superposition matrices representing contributions from SG EMF and GFL current, respectively. The element $\dot{D}^{G}$ is a phasor, with magnitude and angle denoted as $D^{G}$ and $\zeta^{G}$, respectively. Similarly, $\dot{D}^{F} $ is expressed as $D^{F} \angle \zeta^{F}$

For the voltage at the $i^{N}$-th network node, corresponding to the $i^{N}$-th element in (\ref{eq:USS}), we have:
\begin{equation}\label{eq:USS_e}
    \dot{U}_{i^{N}}^{N} = \sum_{\forall i^{G} \in N^{G}} \dot{D}_{i^{N},i^{G}}^{G} \dot{E}_{i^{G}}^{G} + \sum_{\forall i^{F} \in N^{F}} \dot{D}_{i^{N},i^{F}}^{F} \dot{I}_{i^{F}}^{F},
\end{equation}
where $\dot{D}_{i^{N},i^{G}}^{G}$ is the $i^{N}$-th row and $i^{G}$-th column of $\boldsymbol{\dot{D}}^{G}$, and a similar definition applies to $\dot{D}_{i^{N},i^{F}}^{F}$; the indices $i^{G}$ and $i^{F}$ correspond to SGs and GFLs, respectively.

Equations (\ref{eq:USS}) and (\ref{eq:USS_e}) describe the node voltage superposition resulting from the state variables of SGs and GFLs. Since the node voltage, SG, and GFL state variables are all represented as phasors, the node frequency superposition relationship is implicitly embedded within (\ref{eq:USS}) or (\ref{eq:USS_e}). A method described in \cite{JLiu23} can be referenced to extract this relationship. To begin, a general phasor $\dot{F}$, with instantaneous frequency $\omega$, which can represent $\dot{U}^{N} = U^{N} \angle \theta^{N}$, $\dot{E}^{G} = E^{G} \angle \delta$, or $\dot{I}^{F}=I^{F} \angle \theta^{F,I}$ with their respective frequencies $\omega^{N}$, $\omega^{G}$, and $\omega^{F}$, is expressed as:
\begin{subequations}\label{eq:GPhasor}
\begin{equation}\label{eq:GPhasor_F}
    \dot{F} = F \angle \theta,
\end{equation}
\begin{equation}\label{eq:GPhasor_ang}
    \theta = \omega^{0} \int \Delta \omega \mathrm{d} t + \theta^{init},
\end{equation}
\end{subequations}
where the superscript $\{\cdot\}^{init}$ denotes the initial value.

To extract the frequency from the amplitude and angle dynamics in (\ref{eq:GPhasor}), the time derivative of (\ref{eq:GPhasor_F}) is calculated by differentiating both the amplitude $F$ and the angle $\theta$ terms. This yields:
\begin{equation}\label{eq:PhasorDiv}
    \frac{\mathrm{d} \dot{X}}{\mathrm{d} t} = e^{j \theta} \frac{\mathrm{d} X}{\mathrm{d} t} + j \omega^{0} X e^{j \theta} \Delta \omega.
\end{equation}
For the phasors $\dot{U}^{N}$, $\dot{E}^{G}$, and $\dot{I}^{F}$, their time derivatives can be expressed based on (\ref{eq:PhasorDiv}) as follows:
\begin{subequations}\label{eq:PhasorDivEUI}
\begin{equation}\label{eq:PhasorDivEUI_U}
    \frac{\mathrm{d} \dot{U}^{N}}{\mathrm{d} t} = e^{j \theta^{N}} \frac{\mathrm{d} U^{N}}{\mathrm{d} t} + j \omega^{0} U^{N} e^{j \theta^{N}} \Delta \omega^{N},
\end{equation}
\begin{equation}\label{eq:PhasorDivEUI_E}
    \frac{\mathrm{d} \dot{E}^{G}}{\mathrm{d} t} = j \omega^{0} E^{G} e^{j \delta^{G}} \Delta \omega^{G},
\end{equation}
\begin{equation}\label{eq:PhasorDivEUI_I}
    \frac{\mathrm{d} \dot{I}^{F}}{\mathrm{d} t} = e^{j \theta^{F,I}} \frac{\mathrm{d} I^{F}}{\mathrm{d} t} + j \omega^{0} I^{F} e^{j \theta^{F,I}} \Delta \omega^{F}.
\end{equation}
\end{subequations}
It should be noted that the time derivative of the amplitude part of $\dot{E}^{G}$ is zero, as it is constant.

By substituting (\ref{eq:PhasorDivEUI}) into (\ref{eq:USS_e}) and expressing $\dot{D}^{G}$ and $\dot{D}^{F}$ in their phasor forms, then applying a time derivative to both sides of the equation, we obtain:
\begin{equation}\label{eq:FreqOrg}
\begin{aligned}
    & e^{j \theta_{i^{N}}^{N}} \frac{\mathrm{d} U_{i^{N}}^{N}}{\mathrm{d} t} + j \omega^{0} U_{i^{N}}^{N} e^{j \theta_{i^{N}}^{N}} \Delta \omega_{i^{N}}^{N} = \\
    & \sum_{\forall i^{G} \in N^{G}} D_{i^{N},i^{G}}^{N,G} e^{j \zeta_{i^{N},i^{G}}^{G}} \left( j \omega^{0} E_{i^{G}}^{G} e^{j \delta_{i^{G}}^{G}} \Delta \omega_{i^{G}}^{G} \right) \\
    & + \! \! \! \sum_{\forall i^{F} \in N^{F}} \! \! \! \! \! D_{i^{N},i^{F}}^{N,F} e^{j \zeta_{i^{N},i^{F}}^{F}} \! \! \left(\! e^{j \theta_{i^{F}}^{F,I}} \frac{\mathrm{d} I_{i^{F}}^{F}}{\mathrm{d} t} \! + \! j \omega^{0} I_{i^{F}}^{F} e^{j \theta_{i^{F}}^{F,I}} \Delta \omega_{i^{F}}^{F} \! \! \right). \! \! \! \! \! \! \!
\end{aligned}
\end{equation}
To simplify and separate the real and imaginary parts, both sides of the equation are rotated by the angle $\theta_{i^{N}}^{N}$ and all angle-related components are then combined:
\begin{equation}\label{eq:FreqDiv}
\begin{aligned}
    & \frac{\mathrm{d} U_{i^{N}}^{N}}{\mathrm{d} t} + j \omega^{0} U_{i^{N}}^{N} \Delta \omega_{i^{N}}^{N} = \\
    & \sum_{\forall i^{G} \in N^{G}} j \omega^{0} E_{i^{G}}^{G} D_{i^{N},i^{G}}^{N,G} e^{j \left( \delta_{i^{G}}^{G} - \theta_{i^{N}}^{N} + \zeta_{i^{N},i^{G}}^{G} \right)} \Delta \omega_{i^{G}}^{G} \\
    & + \sum_{\forall i^{F} \in N^{F}}  \left( D_{i^{N},i^{F}}^{N,F} e^{j \left(\theta_{i^{F}}^{F,I} - \theta_{i^{N}}^{N} + \zeta_{i^{N},i^{F}}^{F}\right)} \frac{\mathrm{d} I_{i^{F}}^{F}}{\mathrm{d} t} \right. \\
    & \left. + j \omega^{0} I_{i^{F}}^{F} D_{i^{N},i^{F}}^{N,F} e^{j \left(\theta_{i^{F}}^{F,I} - \theta_{i^{N}}^{N} + \zeta_{i^{N},i^{F}}^{F}\right)} \Delta \omega_{i^{F}}^{F} \right).
\end{aligned}
\end{equation}
Next, by extracting the imaginary part, which contains the node frequency $\omega^{N}$, we have:
\begin{equation}\label{eq:FreqImg}
\begin{aligned}
    & j \omega^{0} U_{i^{N}}^{N} \Delta \omega_{i^{N}}^{N} = \\
    & \sum_{\forall i^{G} \in N^{G}} j \omega^{0} E_{i^{G}}^{G} D_{i^{N},i^{G}}^{N,G} \cos \left( \delta_{i^{G}}^{G} - \theta_{i^{N}}^{N} + \zeta_{i^{N},i^{G}}^{G} \right) \Delta \omega_{i^{G}}^{G} \! \! \\
    & + \sum_{\forall i^{F} \in N^{F}}  \left( j D_{i^{N},i^{F}}^{N,F} \sin \left(\theta_{i^{F}}^{F,I} - \theta_{i^{N}}^{N} + \zeta_{i^{N},i^{F}}^{F}\right) \frac{\mathrm{d} I_{i^{F}}^{F}}{\mathrm{d} t} \right. \! \! \\
    & \left. + j \omega^{0} I_{i^{F}}^{F} D_{i^{N},i^{F}}^{N,F} \cos \left(\theta_{i^{F}}^{F,I} - \theta_{i^{N}}^{N} + \zeta_{i^{N},i^{F}}^{F}\right) \Delta \omega_{i^{F}}^{F} \right). \! \!
\end{aligned}
\end{equation}

Dividing both sides by $j \omega^{0} U_{i^{N}}^{N}$, the extended FD formula for node frequency is obtained:
\begin{subequations}\label{eq:FDF}
\begin{equation}\label{eq:FDF_eq}
\begin{aligned}
    \Delta \omega_{i^{N}}^{N} = & \sum_{\forall i^{G} \in N^{G}} A_{i^{N},i^{G}}^{N,G} \Delta \omega_{i^{G}}^{G} \\
    & + \sum_{\forall i^{F} \in N^{F}}  \left( A_{i^{N},i^{F}}^{N,F,Ph} \Delta \omega_{i^{F}}^{F} + A_{i^{N},i^{F}}^{N,F,Am} \frac{\mathrm{d} I_{i^{F}}^{F}}{\mathrm{d} t} \right), \! \! \! \! \! \!
\end{aligned}
\end{equation}
\begin{equation}\label{eq:FDF_AG}
    A_{i^{N},i^{G}}^{N,G} = \frac{E_{i^{G}}^{G}}{U_{i^{N}}^{N}} D_{i^{N},i^{G}}^{N,G} \cos \left( \delta_{i^{G}}^{G} - \theta_{i^{N}}^{N} + \zeta_{i^{N},i^{G}}^{G} \right),
\end{equation}
\begin{equation}\label{eq:FDF_AFph}
    A_{i^{N},i^{F}}^{N,F,Ph} = \frac{I_{i^{F}}^{F}}{U_{i^{N}}^{N}} D_{i^{N},i^{F}}^{N,F} \cos \left(\theta_{i^{F}}^{F,I} - \theta_{i^{N}}^{N} + \zeta_{i^{N},i^{F}}^{F}\right).
\end{equation}
\begin{equation}\label{eq:FDF_AFam}
    A_{i^{N},i^{F}}^{N,F,Am} = \frac{1}{\omega^{0} U_{i^{N}}^{N}} D_{i^{N},i^{F}}^{N,F} \sin \left(\theta_{i^{F}}^{F,I} - \theta_{i^{N}}^{N} + \zeta_{i^{N},i^{F}}^{F}\right), \! \!
\end{equation}
\end{subequations}
where $A_{i^{N},i^{G}}^{N,G}$, $A_{i^{N},i^{F}}^{N,F,Ph}$, and $A_{i^{N},i^{F}}^{N,F,Am}$ are the frequency superposition coefficients for node frequency.

In vector form, similar to the traditional FD formula (\ref{eq:TdFDF}), the extended FD formula is given by:
\begin{equation}\label{eq:FDF_Mtx}
    \Delta \boldsymbol{\omega}^{N} = \boldsymbol{A}^{N,G} \Delta \boldsymbol{\omega}^{G} + \boldsymbol{A}^{N,F,Ph} \Delta \boldsymbol{\omega}^{F} + \boldsymbol{A}^{N,F,Am} \frac{\mathrm{d} \boldsymbol{I}^{F}}{\mathrm{d} t},
\end{equation}
where the frequency superposition matrices satisfy $\boldsymbol{A}^{N,G} \in \mathbb{R}_{N^{N} \times N^{G}}$, $\boldsymbol{A}^{N,F,Ph} \in \mathbb{R}_{N^{N} \times N^{F}}$, and $\boldsymbol{A}^{N,F,Am} \in \mathbb{R}_{N^{N} \times N^{F}}$, with their elements defined in (\ref{eq:FDF_AG})-(\ref{eq:FDF_AFph}).

\subsection{Extended FD Formula for Branch Frequency\label{SubSec:ExtFDFI}}

Unlike traditional systems, where the state variable is the SG EMF, the interface state variable in systems with GFLs also includes the GFL current. Consequently, the branch frequency, which is an important index for system dynamics \cite{IEEEC37}, exhibits spatial variations across the system.

\begin{figure}[!t]
	\centering
	\includegraphics[width=0.3\textwidth]{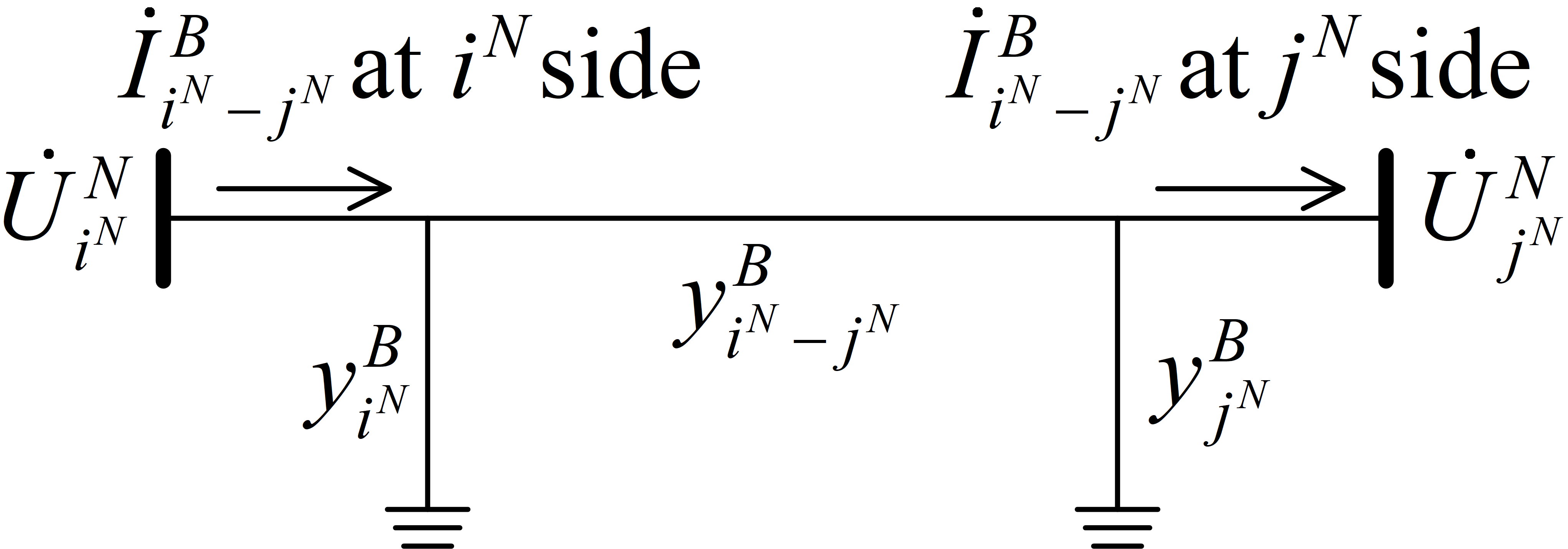}
	\caption{Current phasor in power network.}
	\label{fig:LineTopo}
\end{figure}

A branch representing a transmission line or transformer is illustrated in Fig. \ref{fig:LineTopo}, where the branch connects the $i^{N}$-th node and the $j^{N}$-th node. The current phasors at the start and end terminals are denoted as $\dot{I}_{i^{N}-j^{N}}^{B}$. The admittance between the two terminals is represented by $\dot{y}_{i^{N}-j^{N}}^{B}$, while the admittances to the ground are $\dot{y}_{i^{N}}^{B}$ and $\dot{y}_{j^{N}}^{B}$. The current phasor can be expressed in terms of the terminal voltages as follows, with separate expressions for the start and end terminals in (\ref{eq:CurtDU_sta}) and (\ref{eq:CurtDU_end}):
\begin{subequations}\label{eq:CurtDU}
\begin{equation}\label{eq:CurtDU_sta}
    \dot{I}_{i^{N}-j^{N}}^{B} = \left(\dot{y}_{i^{N}-j^{N}}^{B} + \dot{y}_{i^{N}}^{B}\right) \dot{U}_{i^{N}}^{N} - \dot{y}_{i^{N}-j^{N}}^{B} \dot{U}_{j^{N}}^{N},
\end{equation}
\begin{equation}\label{eq:CurtDU_end}
    \dot{I}_{i^{N}-j^{N}}^{B} = \dot{y}_{i^{N}-j^{N}}^{B} \dot{U}_{i^{N}}^{N} - \left(\dot{y}_{i^{N}-j^{N}}^{B} + \dot{y}_{i^{N}}^{B}\right) \dot{U}_{j^{N}}^{N}.
\end{equation}
\end{subequations}

Following similar steps to those used in deriving the node frequency in Section \ref{SubSec:ExtFDFV}, substituting (\ref{eq:USS_e}) into (\ref{eq:CurtDU}) allows the branch current to be represented in terms of the state variables of SGs and GFLs:
\begin{subequations}\label{eq:CurrSup}
\begin{equation}\label{eq:CurrSup_eq}
    \dot{I}_{i^{N}-j^{N}}^{B} = \sum_{\forall i^{G} \in N^{G}} \dot{D}_{i^{N}-j^{N},i^{G}}^{B,G} \dot{E}_{i^{G}}^{G} + \sum_{\forall i^{F} \in N^{F}} \dot{D}_{i^{N}-j^{N},i^{F}}^{B,F} \dot{I}_{i^{F}}^{F},
\end{equation}
\begin{equation}\label{eq:CurrSup_DTG_sta}
    \dot{D}_{i^{N}-j^{N},i^{G}}^{B,G} = \left(\dot{y}_{i^{N}-j^{N}}^{B} + \dot{y}_{i^{N}}^{B}\right) \dot{D}_{i^{N},i^{G}}^{G} - \dot{y}_{i^{N}-j^{N}}^{B} \dot{D}_{j^{N},i^{G}}^{G},
\end{equation}
\begin{equation}\label{eq:CurrSup_DTF_sta}
    \dot{D}_{i^{N}-j^{N},i^{F}}^{B,F} = \left(\dot{y}_{i^{N}-j^{N}}^{B} + \dot{y}_{i^{N}}^{B}\right) \dot{D}_{i^{N},i^{F}}^{F} - \dot{y}_{i^{N}-j^{N}}^{B} \dot{D}_{j^{N},i^{F}}^{F},
\end{equation}
\begin{equation}\label{eq:CurrSup_DTG_end}
    \dot{D}_{i^{N}-j^{N},i^{G}}^{B,G} = \dot{y}_{i^{N}-j^{N}}^{B} \dot{D}_{i^{N},i^{G}}^{G} - \left(\dot{y}_{i^{N}-j^{N}}^{B} + \dot{y}_{i^{N}}^{B}\right) \dot{D}_{j^{N},i^{G}}^{G},
\end{equation}
\begin{equation}\label{eq:CurrSup_DTF_end}
    \dot{D}_{i^{N}-j^{N},i^{F}}^{B,F} = \dot{y}_{i^{N}-j^{N}}^{B} \dot{D}_{i^{N},i^{F}}^{F} - \left(\dot{y}_{i^{N}-j^{N}}^{B} + \dot{y}_{i^{N}}^{B}\right) \dot{D}_{j^{N},i^{F}}^{F},
\end{equation}
\end{subequations}
where $\dot{D}_{i^{N}-j^{N},i^{G}}^{B,G} \in \mathbb{R}_{N^{B} \times N^{G}}$ and $\dot{D}_{i^{N}-j^{N},i^{F}}^{B,F} \in \mathbb{R}_{N^{B} \times N^{F}}$ are the branch current phasor superposition coefficients corresponding to SGs and GFLs, respectively; $N^{B}$ is the total number of branches in the system. Note that (\ref{eq:CurrSup_DTG_sta}) and (\ref{eq:CurrSup_DTF_sta}) are for the current at the start terminal, while (\ref{eq:CurrSup_DTG_end}) and (\ref{eq:CurrSup_DTF_end}) are for the end terminal.

Next, the phasors are expressed in their amplitude and angle forms as follows: $\dot{I}_{i^{N}-j^{N}}^{B} = I_{i^{N}-j^{N}}^{B} \angle \theta_{i^{N}-j^{N}}^{B}$, $\dot{D}_{i^{N}-j^{N},i^{G}}^{B,G} = D_{i^{N}-j^{N},i^{G}}^{B,G} \angle \zeta_{i^{N}-j^{N},i^{G}}^{B,G}$, $\dot{D}_{i^{N}-j^{N},i^{F}}^{B,F} = D_{i^{N}-j^{N},i^{F}}^{B,F} \angle \zeta_{i^{N}-j^{N},i^{F}}^{B,F}$. An equation similar to (\ref{eq:FDF}), representing the branch frequency superposition, can then be formulated as:
\begin{subequations}\label{eq:FDFT}
\begin{equation}\label{eq:FDFT_eq}
\begin{aligned}
    \Delta \omega_{i^{N}-j^{N}}^{B} = & \sum_{\forall i^{G} \in N^{G}} A_{i^{N}-j^{N},i^{G}}^{B,G} \Delta \omega_{i^{G}}^{G} + \! \! \sum_{\forall i^{F} \in N^{F}} \! \!  \times \\
    & \left( A_{i^{N}-j^{N},i^{F}}^{B,F,Ph} \Delta \omega_{i^{F}}^{F} + A_{i^{N}-j^{N},i^{F}}^{B,F,Am} \frac{\mathrm{d} I_{i^{F}}^{F}}{\mathrm{d} t} \right) \! ,
\end{aligned}
\end{equation}
\begin{equation}\label{eq:FDFT_AG}
\begin{aligned}
    A_{i^{N}-j^{N},i^{G}}^{B,G} = & \frac{E_{i^{G}}^{G}}{I_{i^{N}-j^{N}}^{B}} D_{i^{N}-j^{N},i^{G}}^{B,G} \\
    & \times \cos \left( \delta_{i^{G}}^{G} - \theta_{i^{N}-j^{N}}^{B} + \zeta_{i^{N}-j^{N},i^{G}}^{B,G} \right),
\end{aligned}
\end{equation}
\begin{equation}\label{eq:FDFT_AFph}
\begin{aligned}
    A_{i^{N}-j^{N},i^{F}}^{B,F,Ph} = & \frac{I_{i^{F}}^{F}}{I_{i^{N}-j^{N}}^{B}} D_{i^{N}-j^{N},i^{F}}^{B,F} \\
    & \times \cos \left(\theta_{i^{F}}^{F,I} - \theta_{i^{N}-j^{N}}^{B} + \zeta_{i^{N}-j^{N},i^{F}}^{B,F} \right),
\end{aligned}
\end{equation}
\begin{equation}\label{eq:FDFT_AFam}
\begin{aligned}
    A_{i^{N}-j^{N},i^{F}}^{B,F,Am} = & \frac{1}{\omega^{0} I_{i^{N}-j^{N}}^{B}} D_{i^{N}-j^{N},i^{F}}^{B,F} \\
    & \times \sin \left(\theta_{i^{F}}^{F,I} - \theta_{i^{N}-j^{N}}^{B} + \zeta_{i^{N}-j^{N},i^{F}}^{B,F} \right),
\end{aligned}
\end{equation}
\end{subequations}
where $A_{i^{N}-j^{N},i^{G}}^{B,G}$, $A_{i^{N}-j^{N},i^{F}}^{B,F,Ph}$, and $A_{i^{N}-j^{N},i^{F}}^{B,F,Am}$ are the frequency superposition coefficients for branch frequency.

Additionally, a vectorized form can include all branches in the system:
\begin{equation}\label{eq:FDFT_Mtx}
    \Delta \boldsymbol{\omega}^{B} = \boldsymbol{A}^{B,G} \Delta \boldsymbol{\omega}^{G} + \boldsymbol{A}^{B,F,Ph} \Delta \boldsymbol{\omega}^{F} + \boldsymbol{A}^{B,F,Am} \frac{\mathrm{d} \boldsymbol{I}^{F}}{\mathrm{d} t},
\end{equation}
where matrices $\boldsymbol{A}^{B,G} \in \mathbb{R}_{N^{B} \times N^{G}}$, $\boldsymbol{A}^{B,F,Ph} \in \mathbb{R}_{N^{B} \times N^{F}}$, and $\boldsymbol{A}^{B,F,Am} \in \mathbb{R}_{N^{B} \times N^{F}}$, with their elements defined in (\ref{eq:FDFT_AG})-(\ref{eq:FDFT_AFam}).

\section{Effect of GFLs on Frequency Spatial Variation\label{Sec:GflEffc}}

The characteristics of GFL frequency source are emphasized in Section \ref{SubSec:SourImpt}. The contributions of SG and GFL frequencies to node frequency are analyzed in Section \ref{SubSec:DegrVolt}. The contributions on branch frequency are discussed in Section \ref{SubSec:DegrCurr}.

\subsection{Sources of Frequency\label{SubSec:SourImpt}}

For the original FD formula in (\ref{eq:TdFDF}), the node frequency is determined solely by the SG rotor frequency $\omega^{G}$. However, in the extended FD formula given by (\ref{eq:FDF_Mtx}) and (\ref{eq:FDFT_Mtx}), the node frequency or branch frequency includes additional contributions from the GFL side:
\begin{itemize}
    \item \textbf{Equivalent frequency $\boldsymbol{\omega^{F}}$}: As described in (\ref{eq:SSDynW_w}), this source represents the rate of change of the angle part of the interface state variable, $\theta^{F,I}$, and includes two components. The $\omega^{F,Pll}$ component originates from the state variable of the PLL angle, $\theta^{PLL}$. Its dynamics are influenced by the PLL, d-axis power control, and DC boost. The $\omega^{F,Id}$ component originates from the state variable of the GFL d-axis current, $i_{d}^{P}$, governed by the d-axis power control and DC boost.
    \item \textbf{Equivalent frequency from current amplitude $\boldsymbol{\mathrm{d} I^{F} / \mathrm{d} t}$}: The rate of change of the amplitude part of the state variable, $I^{F}$, also contributes to the network frequency. However, as shown later, the contribution from this term is minimal.
\end{itemize}

\subsection{Superposition Matrices for Node Frequency\label{SubSec:DegrVolt}}

In (\ref{eq:FDF_Mtx}), three sources of frequency, that is, the $\omega^{G}$ of SGs and $\omega^{F}$ and $\mathrm{d} I^{F} / \mathrm{d} t$ of GFLs, have their own superposition matrices $\boldsymbol{A}^{N,G}$, $\boldsymbol{A}^{N,F,Ph}$, and $\boldsymbol{A}^{N,F,Am}$ determining their impact degree on node frequency. Closely digging into their detailed expressions as provided in (\ref{eq:FDF_AG})-(\ref{eq:FDF_AFam}), two types of determinants are included.

The first is \textbf{network topology/parameters-related} determinants. These determinants are the node voltage superposition coefficients $D^{G} \angle \zeta^{G}$ for SGs and $D^{F} \angle \zeta^{F}$ for GFLs. According to (\ref{eq:USS_DG}) and (\ref{eq:USS_DF}), $D^{G} \angle \zeta^{G}$ and $D^{F} \angle \zeta^{F}$ are derived from the system connection matrices $\boldsymbol{\dot{Y}}^{eq,NG}$, $\boldsymbol{\dot{T}}^{eq,NF}$, and $\boldsymbol{\dot{Y}}^{eq,NN}$ given in (\ref{eq:CoefYTZ_NG})-(\ref{eq:CoefYTZ_NN}). The amplitude and angle components of these determinants are analyzed separately:
\begin{itemize}
    \item \textbf{Amplitudes $\boldsymbol{D^{G}}$ and $\boldsymbol{D^{F}}$}: Since they are derived from the system connection matrix, their values depend on the electrical distance between the observation node and the contributing generator. This means the node frequency is affected mainly by its nearby generators. This conclusion is consistent with the traditional FD for SGs \cite{FMilano17}. However, the values of $D^{G}$ and $D^{F}$ are quite different if we look closely at the $\boldsymbol{\dot{Y}}^{eq,NG}$ and $\boldsymbol{\dot{T}}^{eq,NF}$ building them, as given in (\ref{eq:USS_DG}) and (\ref{eq:USS_DF}), respectively. $\boldsymbol{\dot{T}}^{eq,NF}$ is much smaller than $\boldsymbol{\dot{Y}}^{eq,NG}$ since it does not come with a summation term like the original admittance matrix, such as $\boldsymbol{\dot{Y}}^{NG}$ in $\boldsymbol{\dot{Y}}^{eq,NG}$. Thus, the value of $D^{F}$ is much smaller than $D^{G}$, leading to the conclusion that the GFL contribution to node frequency is much smaller than that of SGs.
    \item \textbf{Angles $\boldsymbol{\zeta^{G}}$ and $\boldsymbol{\zeta^{F}}$}: Their characteristics are different also because of the different forms of $\boldsymbol{\dot{Y}}^{eq,NG}$ and $\boldsymbol{\dot{T}}^{eq,NF}$. For SGs, $\zeta^{G}$ is close to $0$ due to the predominance of reactance over resistance in the system, consistent with the findings in \cite{FMilano17}. For GFLs, due to the even times multiplication of the system admittance matrix making the angle of $\boldsymbol{\dot{T}}^{eq,NF}$ shift by $\pi / 2$ from the system admittance matrix, $\zeta^{F}$ is approximately $\pi / 2$. The contribution degree of angles $\boldsymbol{\zeta^{G}}$ and $\boldsymbol{\zeta^{F}}$ to the frequency superposition matrices $\boldsymbol{A}^{N,G}$, $\boldsymbol{A}^{N,F,Ph}$, and $\boldsymbol{A}^{N,F,Am}$ should be analyzed together with the other angle terms discussed later.
\end{itemize}

The other is \textbf{power flow-related} determinants. These determinants involve $E^{G} / U^{N}$ and $\delta^{G} - \theta^{N}$ for measuring source $\omega^{G}$ of the SG, $I^{F} / U^{N}$ and $\theta^{F,I} - \theta^{N}$ for GFL $\omega^{F}$, and $1 / \left(\omega^{0} U^{N}\right)$ and $\theta^{F,I} - \theta^{N}$ for the GFL $\mathrm{d} I^{F} / \mathrm{d} t$. These terms are determined by the system power flow. The amplitude and angle components are analyzed as follows:
\begin{itemize}
    \item \textbf{Amplitudes $\boldsymbol{E^{G} / U^{N}}$, $\boldsymbol{I^{F} / U^{N}}$, and $\boldsymbol{1 / \left(\omega^{0} U^{N}\right)}$}: $E^{G} / U^{N}$ and $I^{F} / U^{N}$ for $\omega^{G}$ and $\omega^{F}$ are close to $1$, which does not affect the superposition relations $\boldsymbol{A}^{N,G}$ and $\boldsymbol{A}^{N,F,Ph}$. However, $1 / \left(\omega^{0} U^{N}\right)$ for $\mathrm{d} I^{F} / \mathrm{d} t$ is much smaller due to the division by $\omega^{0}$, indicating a minimal contribution to the node frequency through the source $\mathrm{d} I^{F} / \mathrm{d} t$ of the GFL.
    \item \textbf{Angles $\boldsymbol{\delta^{G} - \theta^{N}}$ and $\boldsymbol{\theta^{F,I} - \theta^{N}}$}: These angles are influenced by the active power flow. A higher generator power output typically results in a larger angle difference between the generator and node voltage phasors, leading to a higher contribution from this generator to the node frequency after combining the angles $\boldsymbol{\zeta^{G}}$ and $\boldsymbol{\zeta^{F}}$.
\end{itemize}

In addition, the power flow-related determinants are time-varying, increasing the complexity of applying the FD formula online since more variables must be measured in real-time to determine system frequency correlations. However, as the system operating condition remains relatively stable over short periods, these terms can be considered constant during such intervals. The frequency superposition coefficients in the FD formula (\ref{eq:FDF_Mtx}) can then be fixed as:
\begin{subequations}\label{eq:simFDF}
\begin{equation}\label{eq:simFDF_AG}
    A_{i^{N},i^{G}}^{N,G0} = \frac{E_{i^{G}}^{G0}}{U_{i^{N}}^{N0}} D_{i^{N},i^{G}}^{N,G} \cos \left( \delta_{i^{G}}^{G0} - \theta_{i^{N}}^{N0} + \zeta_{i^{N},i^{G}}^{G} \right)
\end{equation}
\begin{equation}\label{eq:simFDF_AFph}
    A_{i^{N},i^{F}}^{N,F,Ph0} = \frac{I_{i^{F}}^{F0}}{U_{i^{N}}^{N0}} D_{i^{N},i^{F}}^{N,F} \cos \left(\theta_{i^{F}}^{F,I0} - \theta_{i^{N}}^{N0} + \zeta_{i^{N},i^{F}}^{F}\right)
\end{equation}
\begin{equation}\label{eq:simFDF_AFam}
    A_{i^{N},i^{F}}^{N,F,Am0} = \frac{1}{\omega^{0} U_{i^{N}}^{N0}} D_{i^{N},i^{F}}^{N,F} \sin \left(\theta_{i^{F}}^{F,I0} - \theta_{i^{N}}^{N0} + \zeta_{i^{N},i^{F}}^{F}\right)
\end{equation}
\end{subequations}
where the superscript $\{\cdot\}^{0}$ denotes values at a specific time point during a period.

\subsection{Superposition Matrices for Branch Frequency\label{SubSec:DegrCurr}}

Similar to the case of node frequency, as given in (\ref{eq:FDFT_Mtx}), $\boldsymbol{A}^{B,G}$, $\boldsymbol{A}^{B,F,Ph}$, and $\boldsymbol{A}^{B,F,Am}$ constitute the branch frequency superposition, with their expressions provided in (\ref{eq:FDFT_AG})–(\ref{eq:FDFT_AFam}). There are two types of determinants: system topology/parameter-related and power flow-related.

For the \textbf{system topology/parameter-related} determinants, the conclusions are similar to those for node superposition. The values of the amplitudes $D^{B,G}$ and $D^{B,F}$ depend on the electrical distance between the observed branch location and the contributing generator. The impact from GFLs is relatively small. However, the angles $\zeta^{B,G}$ and $\zeta^{B,F}$ are not close to $0$ but are more variable. This is because the branch admittances $\dot{y}^{B}$ are multiplied with the voltage superposition coefficients $\dot{B,D}^{G}$ or $\dot{B,D}^{F}$ to form the current superposition $\dot{D}^{B,G}$ or $\dot{D}^{B,F}$, as shown in (\ref{eq:CurrSup_DTG_sta})–(\ref{eq:CurrSup_DTF_end}). Consequently, the cosine or sine of these angles in (\ref{eq:FDFT_AG})–(\ref{eq:FDFT_AFam}) can be negative, potentially resulting in negative values for the frequency superposition $\boldsymbol{A}^{B,G}$, $\boldsymbol{A}^{B,F,Ph}$, and $\boldsymbol{A}^{B,F,Am}$.

For the \textbf{power flow-related impacts} determinants, the characteristics of $E^{G} / I^{B}$, $I^{F} / I^{B}$, $1 / \left(\omega^{0} I^{B}\right)$, $\delta^{G} - \theta^{B}$, and $\theta^{F,I} - \theta^{B}$ are similar to those discussed in Section \ref{SubSec:DegrVolt}. These terms can also be evaluated at specific time points, leading to fixed values for $A^{B,G0}$, $A^{B,F,Ph0}$, and $A^{B,F,Am0}$.

\section{Case Studies\label{Sec:CaseIEEE9}}

The proposed extended FD formula is tested on the modified WECC $9$-Bus system, where G3 is replaced by $80 \times 2$MW PMSG wind generators. The system's topology and dynamic parameters are claimed in Part I of this series. The spatial-related parameters, including transmission lines and the transformer (Tsf), are listed in Table 2 in the Appendix \cite{AddDoc}. The power flow data used for the subsequent analysis is also presented in Table 2 in the Appendix \cite{AddDoc}.

Time-domain simulations are conducted using EMT simulations in MATLAB/Simulink. The node or branch frequency is calculated using the angle differential with a short time window of $1/60$ Hz to capture accurate transient patterns.

\subsection{Accuracy Validation\label{SubSec:AccuVali}}

The error index used in the following analysis is the average difference between the EMT-simulated frequency and the calculated frequency, defined as $\int \left|f^{cal}\left(t\right) - f^{emt}\left(t\right)\right| \mathrm{d} t /T $, in mHz, where $T$ is the time window length for error calculation. The node voltage and branch current frequencies are validated separately.

\subsubsection{Node Frequency} 
\begin{figure}[!t]
	\centering
	\includegraphics[width=0.48\textwidth]{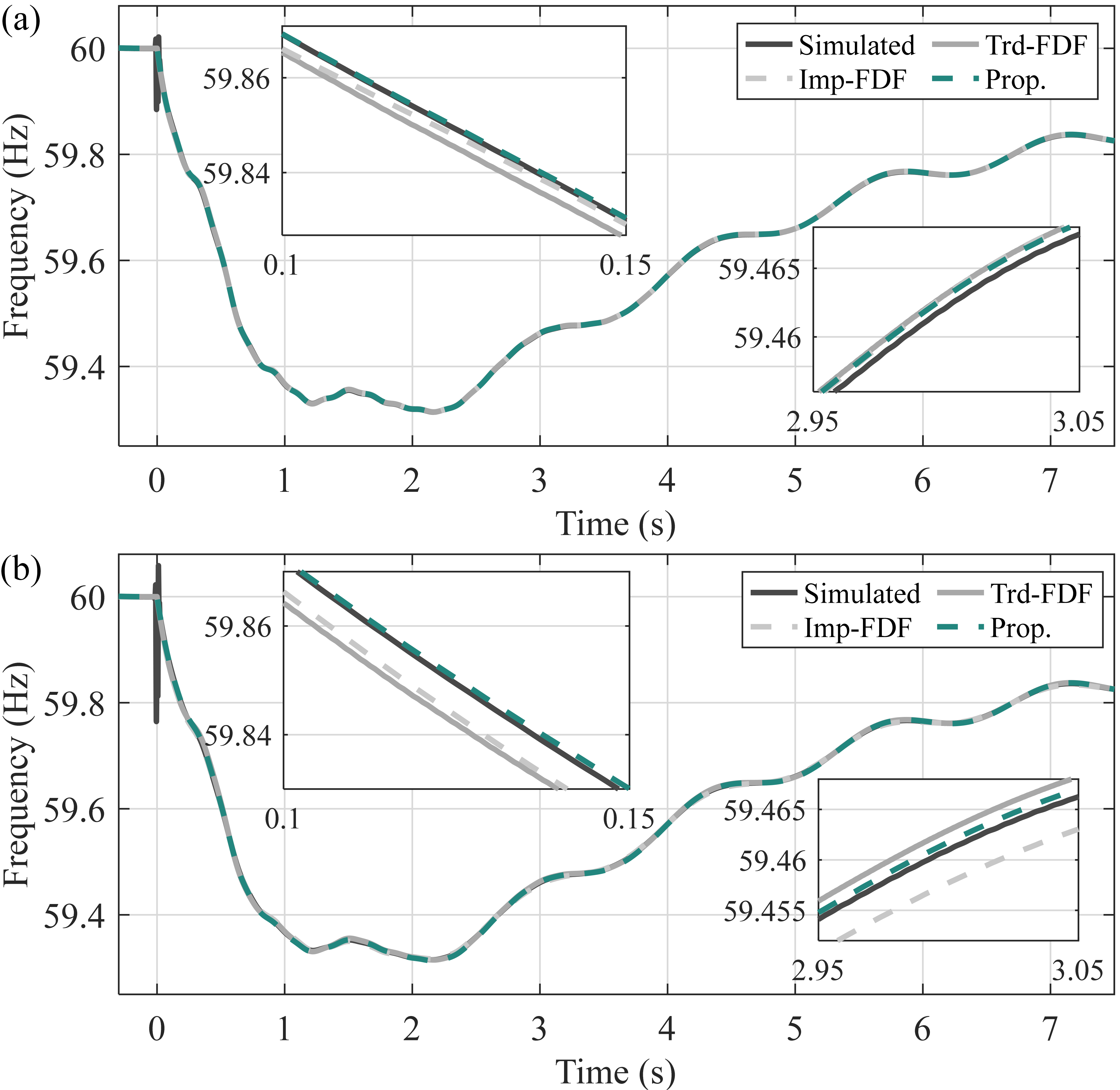}
	\caption{Frequencies of nodes (a) 07 and (b) 11.}
	\label{fig:FreqError}
\end{figure}

A load step disturbance is applied at node $09$. The EMT-simulated frequency dynamics at nodes $07$ and $11$ are shown in Fig. \ref{fig:FreqError}(a) and (b), respectively, with zoomed-in views for certain time periods. Proposed method calculated the frequencies using (\ref{eq:FDF_Mtx}), with superposition matrix considering the dynamics of power flow as in (\ref{eq:FDF_AG})-(\ref{eq:FDF_AFam}), or using fixed parameters as in (\ref{eq:simFDF_AG})-(\ref{eq:simFDF_AFam}). Two other methods are used for comparison:
\begin{itemize}
    \item \textbf{Traditional FD (trd-FD) \cite{FMilano17}}: The node frequency is calculated using (\ref{eq:TdFDF}), where only the SG rotor frequency contributes to the network frequency.
    \item \textbf{Improved FD (imp-FD) \cite{QMa23,QMa24}}: This recent method identifies the contribution of a single GFL to its terminal frequency in \cite{QMa23} and extends it to large-scale systems by combining it with traditional FD in \cite{QMa24}. The time derivatives of the GFL current, PLL frequency, and the derivative of the PLL frequency are included as frequency sources of GFLs, similar to the proposed method.
\end{itemize}

As can be seen in Fig. \ref{fig:FreqError}, although the calculated node frequencies using different methods closely follow the EMT-simulated dynamics over long time scales, significant differences are observed in the zoomed-in plots. For node $07$, as shown in Fig. \ref{fig:FreqError}(a), the ranking of errors is as follows: trd-FD $>$ imp-FD $>$ proposed method, with error indices of $0.91$ mHz, $0.87$ mHz, and $0.41$ mHz, respectively. For node $11$, as shown in Fig. \ref{fig:FreqError}(b), the ranking is imp-FD $>$ trd-FD $>$ proposed method, with error indices of $2.17$ mHz, $1.93$ mHz, and $1.17$ mHz, respectively. The larger error of trd-FD is expected as it neglects the dynamics contributed from GFLs. The imp-FD shows lower error in Fig. \ref{fig:FreqError}(a) due to its inclusion of GFL contributions. However, in Fig. \ref{fig:FreqError}(b), the error surpasses that of trd-FD because imp-FD primarily focuses on single-generator systems \cite{QMa23}, leading to inaccuracies in larger systems, as reflected in case studies in \cite{QMa24}. The proposed method consistently aligns closely with EMT simulations.

\begin{figure}[!t]
	\centering
	\includegraphics[width=0.48\textwidth]{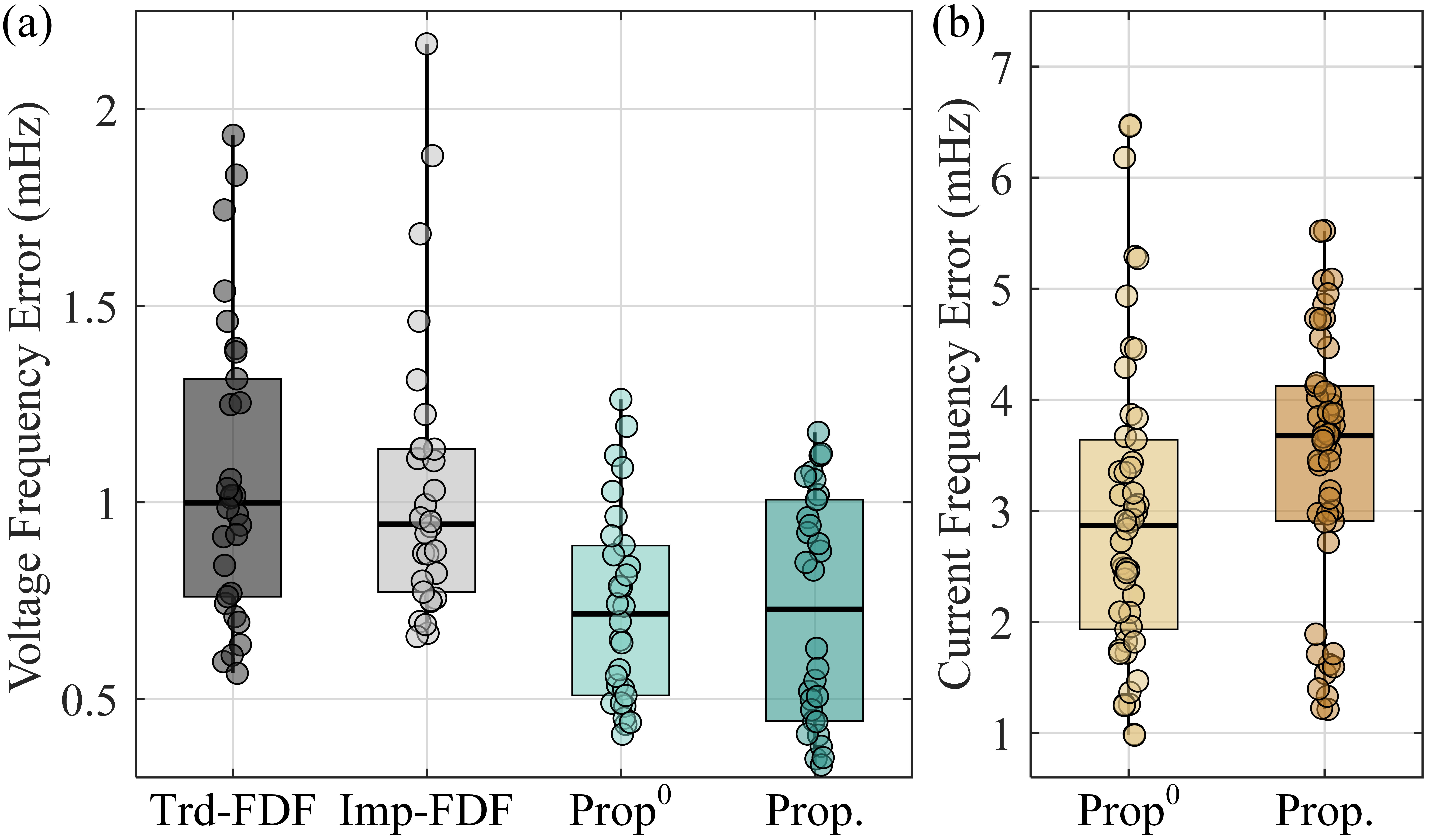}
	\caption{Error indexes of node voltage frequency with different methods.}
	\label{fig:BoxError}
\end{figure}

To evaluate the accuracy under different measuremed nodes and different system conditions, such as varying inertia settings and control parameters, the error indices for the various methods are summarized in Fig. \ref{fig:BoxError}(a). These box plots show the error distribution under each condition, with the median errors summarized in Table \ref{tab:AccrSumm}. The proposed method with fixed parameter settings is also included and is marked as prop$^{0}$. The trd-FD exhibits the largest median error of $1.00$ mHz. The imp-FD has a lower median error of $0.94$ mHz; however, it includes cases with larger errors. The proposed method achieves the lowest errors, with median values of $0.72$ mHz for prop$^{0}$ and $0.73$ mHz for the full proposed method. Compared to the trd-FD, the proposed methods show accuracy improvements of $28.3$\% for prop$^{0}$ and $27.1$\% for the full proposed method, while the imp-FD offers only a $5.4$\% improvement.

\begin{table}[!t]
	\setlength\tabcolsep{4.4pt}
	\setlength{\aboverulesep}{0.0pt}
	\setlength{\belowrulesep}{1.0pt}
	\centering
	\caption{Accuracy Summary}
    \begin{tabular}{ccccccc}
		\toprule
		\specialrule{0em}{0.4pt}{0.4pt}
		\toprule
        \multirow{2}[2]{*}{Index} & \multicolumn{4}{c}{For Voltage} & \multicolumn{2}{c}{For Current} \\
        \cmidrule{2-7}          & Trd-FD & Imp-FD & Prop$^{0}$ & Prop.  & Prop$^{0}$ & Prop. \\
        \midrule
        Median Error (mHz) & 1.00  & 0.94  & 0.72  & 0.73  & 2.87  & 3.68 \\
        Improvement (\%) & 0  & 5.4  & 28.3  & 27.1  & /     & / \\
        \bottomrule
		\specialrule{0em}{0.4pt}{0.4pt}
        \bottomrule
    \end{tabular}
	\label{tab:AccrSumm}
\end{table}

\subsubsection{Branch Frequency} The frequencies of branch current in the system are summarized in Fig. \ref{fig:BoxError}(b) and Table \ref{tab:AccrSumm}. The branch frequency is calculated using (\ref{eq:FDFT_Mtx}) with parameters determined by (\ref{eq:FDFT_AG})-(\ref{eq:FDFT_AFam}). Results using fixed power flow parameters are also presented. As there are no existing methods for branch frequency calculation, the proposed method is not compared to others. The errors shown in Fig. \ref{fig:BoxError}(b) and Table \ref{tab:AccrSumm} indicate acceptable accuracy, with median errors of $2.87$\% and $3.68$\%.

\renewcommand{\thefigure}{4}
\begin{figure*}[!t]
	\centering
	\includegraphics[width=1.0\textwidth]{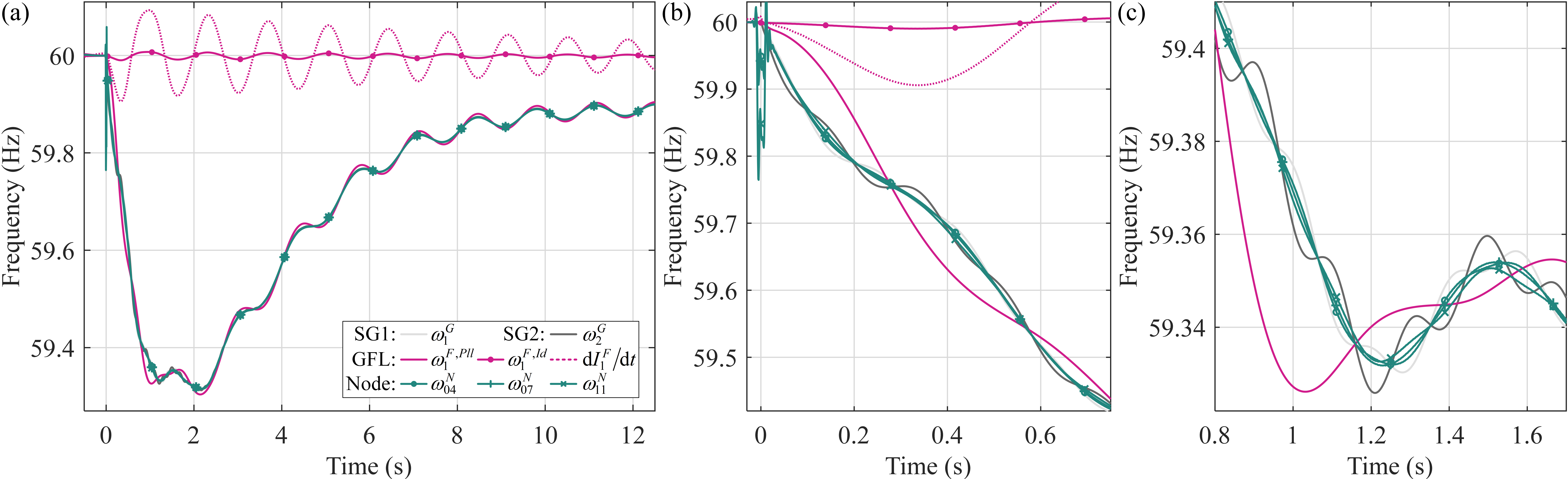}
	\caption{Analysis of GFL impacting node frequency in full-time scale and (b) and (c) zoom-in views.}
	\label{fig:FreqAnaly}
    \begin{equation}\label{eq:CaseHeat_Angl}\tag{23c}
    \boldsymbol{\Psi}_{Ph} \! \! = \! \! 
    \left[
    \renewcommand{\arraycolsep}{1.5pt}
    \begin{matrix}
        \cos \! \left( \delta_{1}^{G} \! \! - \! \theta_{04}^{N} \! \! + \! \zeta_{04,1}^{G} \right) & \cos \! \left( \delta_{2}^{G} \! \! - \! \theta_{04}^{N} \! \! + \! \zeta_{04,2}^{G} \right) & \cos \! \left( \theta_{1}^{F,I} \! \! - \! \theta_{04}^{N} \! \! + \! \zeta_{04,1}^{F} \right) & \sin \! \left( \theta_{1}^{F,I} \! \! - \! \theta_{04}^{N} \! \! + \! \zeta_{04,1}^{F} \right) \\
        \cos \! \left( \delta_{1}^{G} \! \! - \! \theta_{07}^{N} \! \! + \! \zeta_{07,1}^{G} \right) & \cos \! \left( \delta_{2}^{G} \! \! - \! \theta_{07}^{N} \! \! + \! \zeta_{07,2}^{G} \right) & \cos \! \left( \theta_{1}^{F,I} \! \! - \! \theta_{07}^{N} \! \! + \! \zeta_{07,1}^{F} \right) & \sin \! \left( \theta_{1}^{F,I} \! \! - \! \theta_{07}^{N} \! \! + \! \zeta_{07,1}^{F} \right) \\
        \cos \! \left( \delta_{1}^{G} \! \! - \! \theta_{11}^{N} \! \! + \! \zeta_{11,1}^{G} \right) & \cos \! \left( \delta_{2}^{G} \! \! - \! \theta_{11}^{N} \! \! + \! \zeta_{11,2}^{G} \right) & \cos \! \left( \theta_{1}^{F,I} \! \! - \! \theta_{11}^{N} \! \! + \! \zeta_{11,1}^{F} \right) & \sin \! \left( \theta_{1}^{F,I} \! \! - \! \theta_{11}^{N} \! \! + \! \zeta_{11,1}^{F} \right) \\
    \end{matrix}
    \! \right]
    \! \! = \! \! 
    \left[
    \renewcommand{\arraycolsep}{1.5pt}
    \begin{matrix}
        \cellcolor[rgb]{0.000,0.323,0.287} \textcolor{white}{1.00} & \cellcolor[rgb]{0.000,0.323,0.287} \textcolor{white}{1.00} & \cellcolor[rgb]{0.919,0.955,0.951} 0.64 & \cellcolor[rgb]{0.574,0.844,0.801} 0.77 \\
        \cellcolor[rgb]{0.000,0.392,0.360} \textcolor{white}{0.96} & \cellcolor[rgb]{0.000,0.323,0.287} \textcolor{white}{1.00} & \cellcolor[rgb]{0.737,0.904,0.879} 0.72 & \cellcolor[rgb]{0.806,0.926,0.909} 0.69 \\
        \cellcolor[rgb]{0.000,0.374,0.342} \textcolor{white}{0.97} & \cellcolor[rgb]{0.000,0.315,0.278} \textcolor{white}{1.00} & \cellcolor[rgb]{0.927,0.957,0.953} 0.64 & \cellcolor[rgb]{0.559,0.837,0.793} 0.77 \\
    \end{matrix}
    \right] \! \!
    \end{equation}
\end{figure*}

\subsection{Revisit Frequency Spatial Variation\label{SubSec:Revisit}}

The frequency spatial variation with GFLs is analyzed numerically as follows.

\subsubsection{Source of Frequency} The variables plotted in Fig. \ref{fig:FreqAnaly} include the frequencies at nodes $04$, $07$, and $11$, denoted as $\omega_{04}^{N}$, $\omega_{07}^{N}$, and $\omega_{11}^{N}$), respectively; the frequency sources at the SG side, i.e., the rotor frequencies $\omega_{1}^{G}$ and $\omega_{2}^{G}$ corresponding to SG$1$ and SG$2$; and the frequency sources at the GFL side, i.e., the two components $\omega_{1}^{F,Pll}$ and $\omega_{1}^{F,Id}$ of the equivalent frequency $\omega_{1}^{F}$, as well as the equivalent frequency contribution from the current amplitude, $\mathrm{d} I_{1}^{F}/\mathrm{d} t$) Here, the subscript $\{\cdot\}_{1}$ indicates the index of the single GFL in the system. Fig. \ref{fig:FreqAnaly}(a) presents the full-time scale following the disturbance, while Figs. \ref{fig:FreqAnaly}(b) and \ref{fig:FreqAnaly}(c) show zoomed-in views of the initial moment and one second later.

The SG rotor frequencies $\omega_{1}^{G}$ and $\omega_{2}^{G}$ oscillate relative to each other. The PLL component, $\omega_{1}^{F,Pll}$, plays a dominant role in the equivalent frequency $\omega_{1}^{F}$, oscillating around the SG rotor frequency. In contrast, the current component, $\omega_{1}^{F,Id}$, has significantly smaller magnitudes. The equivalent frequency from current amplitude $\mathrm{d} I_{1}^{F}/\mathrm{d} t$ is also with minor magnitude. The dynamics of these state variables are analyzed in detail in Part I of this series. In addition, it is evident that the node frequency is primarily determined by the SGs, with the degree of their impact discussed in the following.

\subsubsection{Superposition Matrices} As discussed in Section \ref{SubSec:DegrVolt}, the frequency superposition matrices determine the degree of impact from various generator frequency sources on different nodes. The determinants of these frequency superposition matrices include factors related to system topology/parameters and power flow, with both categories further separated into amplitude and angle components. Equations (\ref{eq:CaseHeat_AmB})–(\ref{eq:CaseHeat_all}) provide a detailed breakdown of these determinants with numerical values specified: (\ref{eq:CaseHeat_AmB}) represents the amplitude component of system topology/parameters-related determinant; (\ref{eq:CaseHeat_Volt}) represents the amplitude component of the power flow-related determinant; (\ref{eq:CaseHeat_Angl}) combines the angle components of topology/parameter-related and power flow-related; (\ref{eq:CaseHeat_all}) presents the final frequency superposition matrix. The columns of these matrices correspond to the frequency sources: SG rotor frequencies $\omega_{1}^{G}$ and $\omega_{2}^{G}$ for SG$1$ and SG$2$, the GFL equivalent frequency $\omega_{1}^{F}$, and the GFL current amplitude $\mathrm{d} I_{1}^{F}/\mathrm{d} t$. The rows represent the contributions to nodes $04$, $07$, and $11$. The values of each matrix element are visualized using color intensity, with darker shades indicating larger values.

\begin{subequations}\label{eq:CaseHeat}
\begin{equation}\label{eq:CaseHeat_AmB}
    \boldsymbol{\Psi}_{D} \! \! = \! \! 
    \left[
    \renewcommand{\arraycolsep}{1pt}
    \begin{matrix}
        D_{04, 1}^{G} & D_{04, 2}^{G} & D_{04, 1}^{F} & D_{04, 1}^{F} \\
        D_{07, 1}^{G} & D_{07, 2}^{G} & D_{07, 1}^{F} & D_{07, 1}^{F} \\
        D_{11, 1}^{G} & D_{11, 2}^{G} & D_{11, 1}^{F} & D_{11, 1}^{F} \\
    \end{matrix}
    \right]
    \! \! = \! \! 
    \left[
    \renewcommand{\arraycolsep}{1.5pt}
    \begin{matrix}
        \cellcolor[rgb]{0.000,0.315,0.278} \textcolor{white}{0.71} & \cellcolor[rgb]{0.681,0.886,0.853} 0.23 & \cellcolor[rgb]{0.927,0.957,0.953} 0.05 & \cellcolor[rgb]{0.927,0.957,0.953} 0.05 \\
        \cellcolor[rgb]{0.454,0.770,0.724} 0.36 & \cellcolor[rgb]{0.073,0.465,0.435} \textcolor{white}{0.57} & \cellcolor[rgb]{0.902,0.952,0.946} 0.07 & \cellcolor[rgb]{0.902,0.952,0.946} 0.07 \\
        \cellcolor[rgb]{0.252,0.626,0.593} 0.46 & \cellcolor[rgb]{0.346,0.694,0.655} 0.41 & \cellcolor[rgb]{0.774,0.916,0.895} 0.18 & \cellcolor[rgb]{0.774,0.916,0.895} 0.18 \\
    \end{matrix}
    \right]
\end{equation}
\begin{equation}\label{eq:CaseHeat_Volt}
    \boldsymbol{\Psi}_{Am} = 
    \left[
    \renewcommand{\arraycolsep}{1.5pt}
    \begin{matrix}
        \frac{E_{1}^{G}}{U_{04}^{N}} & \frac{E_{2}^{G}}{U_{04}^{N}} & \frac{I_{1}^{F}}{U_{04}^{N}} & \frac{1}{\omega^{0} U_{04}^{N}} \\
        \frac{E_{1}^{G}}{U_{07}^{N}} & \frac{E_{2}^{G}}{U_{07}^{N}} & \frac{I_{1}^{F}}{U_{07}^{N}} & \frac{1}{\omega^{0} U_{07}^{N}} \\
        \frac{E_{1}^{G}}{U_{11}^{N}} & \frac{E_{2}^{G}}{U_{11}^{N}} & \frac{I_{1}^{F}}{U_{11}^{N}} & \frac{1}{\omega^{0} U_{11}^{N}} \\
    \end{matrix}
    \right]
    =
    \left[
    \renewcommand{\arraycolsep}{1.5pt}
    \begin{matrix}
        \cellcolor[rgb]{0.000,0.340,0.305} \textcolor{white}{1.04} & \cellcolor[rgb]{0.000,0.331,0.296} \textcolor{white}{1.05} & \cellcolor[rgb]{0.112,0.504,0.474} \textcolor{white}{0.82} & \cellcolor[rgb]{0.927,0.957,0.953} 0.02 \\
        \cellcolor[rgb]{0.000,0.340,0.305} \textcolor{white}{1.04} & \cellcolor[rgb]{0.000,0.331,0.296} \textcolor{white}{1.05} & \cellcolor[rgb]{0.112,0.504,0.474} \textcolor{white}{0.82} & \cellcolor[rgb]{0.927,0.957,0.953} 0.02 \\
        \cellcolor[rgb]{0.000,0.323,0.287} \textcolor{white}{1.07} & \cellcolor[rgb]{0.000,0.315,0.278} \textcolor{white}{1.08} & \cellcolor[rgb]{0.092,0.485,0.454} \textcolor{white}{0.84} & \cellcolor[rgb]{0.927,0.957,0.953} 0.02 \\
    \end{matrix}
    \right]
\end{equation}
\begin{equation}\label{eq:CaseHeat_all}\tag{23d}
\begin{aligned}
    & \boldsymbol{A} = \boldsymbol{\Psi}_{D} \times \boldsymbol{\Psi}_{Am} \times\boldsymbol{\Psi}_{Ph} \\
    & = \left[
    \renewcommand{\arraycolsep}{1.5pt}
    \begin{matrix}
        A_{04,1}^{G} & A_{04,2}^{G} & A_{04,1}^{F,Ph} & A_{04,1}^{F,Am} \\
        A_{07,1}^{G} & A_{07,2}^{G} & A_{07,1}^{F,Ph} & A_{07,1}^{F,Am} \\
        A_{11,1}^{G} & A_{11,2}^{G} & A_{11,1}^{F,Ph} & A_{11,1}^{F,Am} \\
    \end{matrix}
    \right]
    =
    \left[
    \renewcommand{\arraycolsep}{1.5pt}
    \begin{matrix}
        \cellcolor[rgb]{0.000,0.315,0.278} \textcolor{white}{0.73}
        & \cellcolor[rgb]{0.604,0.857,0.817} 0.24 & \cellcolor[rgb]{0.902,0.952,0.946} 0.03 & \cellcolor[rgb]{0.927,0.957,0.953} 0.001 \\
        \cellcolor[rgb]{0.409,0.739,0.695} 0.36 & \cellcolor[rgb]{0.054,0.447,0.416} \textcolor{white}{0.60} & \cellcolor[rgb]{0.884,0.948,0.940} 0.04 & \cellcolor[rgb]{0.927,0.957,0.953} 0.001 \\
        \cellcolor[rgb]{0.210,0.594,0.563} 0.47 & \cellcolor[rgb]{0.267,0.637,0.603} 0.44 & \cellcolor[rgb]{0.826,0.932,0.918} 0.09 & \cellcolor[rgb]{0.927,0.957,0.953} 0.002 \\
    \end{matrix}
    \right]
\end{aligned}
\end{equation}
\end{subequations}

\renewcommand{\thefigure}{5}
\begin{figure*}[!t]
	\centering
	\includegraphics[width=1.0\textwidth]{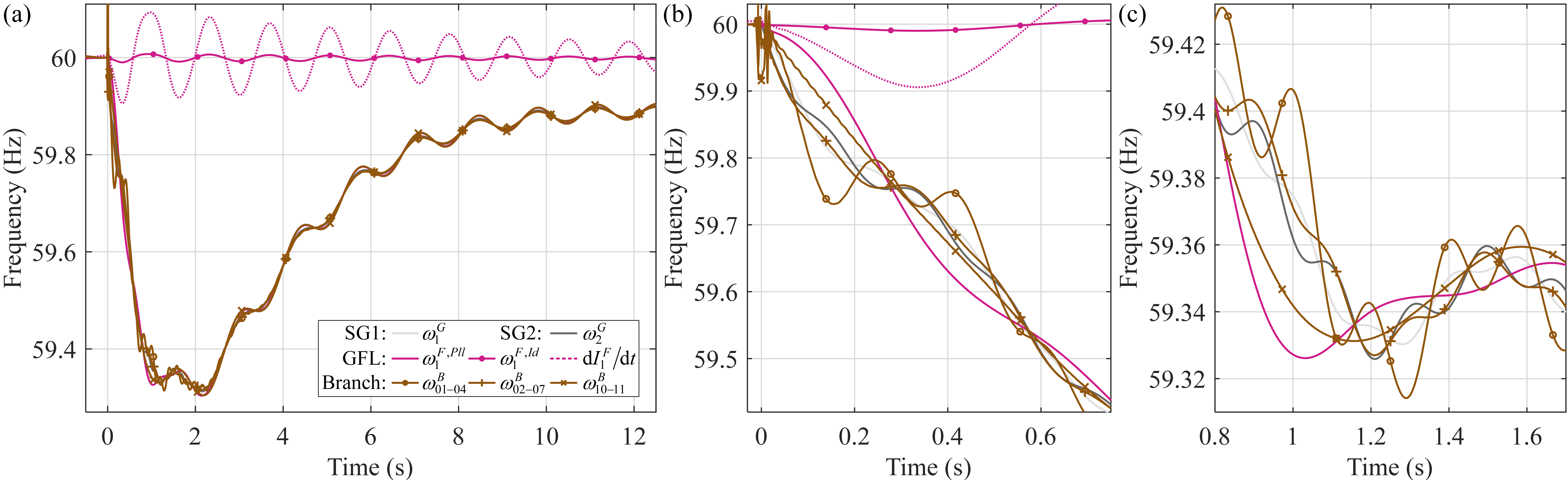}
	\caption{Analysis of GFL impacting branch frequency in full-time scale and (b) and (c) zoom-in views.}
	\label{fig:CurrAnaly}
\end{figure*}

Referring to Section \ref{SubSec:DegrVolt}, the matrices (\ref{eq:CaseHeat_AmB})–(\ref{eq:CaseHeat_Angl}) can be analyzed as follows.
\begin{itemize}
    \item \textbf{(\ref{eq:CaseHeat_AmB})}: The magnitude of each element is influenced by the electrical distance between the source generator and the observation node. For instance, the frequency superposition at node $04$ (first row) shows the largest contribution from generator SG$1$, with $D_{04, 1}^{G} = 0.71$. As discussed in Section \ref{SubSec:DegrVolt}, the contribution from GFLs is significantly smaller due to differences in the calculation of $D^{F}$ compared to $D^{G}$. Even for node $11$, which is closest to the GFL, the superposition coefficient $D_{11, 1}^{F}$ of $\omega^{F}$ or $\mathrm{d} I^{F}/\mathrm{d} t$ is only $0.18$, much smaller than the contribution from SGs.
    \item \textbf{(\ref{eq:CaseHeat_Volt})}: The SG-related superposition coefficients (first two columns) are close to $1$, as expected. The contribution from the GFL equivalent frequency $\omega_{1}^{F}$ (third column) is less than $1$ because the wind farm is not operating at full capacity, meaning $I_{1}^{F0}$ is less than $1$. The contribution from the GFL current amplitude $\mathrm{d} I_{1}^{F}/\mathrm{d} t$ (fourth column) is negligible compared to other sources due to division by $\omega^{0}$.
    \item \textbf{(\ref{eq:CaseHeat_Angl})}: The SG-related coefficients (first two columns) are close to $1$ because: the angle of the voltage superposition coefficient $\zeta^{G}$ between the SGs and the network node is nearly $0$, being dominated by network reactance; the angle difference between the SG EMF and the node voltage, $\delta^{G}-\theta^{N}$, is also close to $0$ due to moderate power transfer. In contrast, the GFL-related impacts (last two columns) are irregular. This irregularity arises because the superposition coefficient $\zeta^{F}$ is close to $\pi/2$, and the angle difference between the current and voltage, $\theta^{F,I}-\theta^{N}$, is not $0$.
\end{itemize}

As a result, the overall frequency superposition matrix in (\ref{eq:CaseHeat_all}) reveals the following patterns: (i) the frequency superposition coefficient is primarily determined by the electrical distance, as evidenced by the larger diagonal elements with darker colors; (ii) the contribution from GFL frequencies is significantly lower than that from SGs due to differences in the formulations of $\dot{D}^{G}$ for SGs and $\dot{D}^{F}$ for GFLs in (\ref{eq:USS}); (iii) the impact of the GFL current amplitude is negligible. Referring back to Fig. \ref{fig:FreqAnaly}, the node frequencies $\omega_{04}^{N}$, $\omega_{07}^{N}$, and $\omega_{11}^{N}$ are predominantly determined by the SG rotor frequencies $\omega_{1}^{G}$ and $\omega_{2}^{G}$, particularly for nodes $04$ and $07$, which are closer to the SGs. The frequency $\omega_{11}^{N}$ at node $11$ is slightly influenced by the GFL equivalent frequency $\omega_{1}^{F}$, especially its PLL-related component, $\omega_{1}^{F,Pll}$, which has a larger magnitude. The contribution from the GFL current amplitude, $\mathrm{d} I^{F}/\mathrm{d} t$, remains negligible.

Another important feature is that the frequency superposition matrix given in (\ref{eq:CaseHeat_all}) is time-varying because it is determined by the system power flow. By reducing the wind speed to decrease the GFL wind farm input from original $0.479$ pu to $0.380$ pu, the matrices (\ref{eq:CaseHeat_Volt}), (\ref{eq:CaseHeat_Angl}), and (\ref{eq:CaseHeat_all}) are updated to (\ref{eq:HeatOprt_Volt}), (\ref{eq:HeatOprt_Angl}), and (\ref{eq:HeatOprt_all}). For the amplitude part of the power flow-related determinant in (\ref{eq:CaseHeat_Volt}), the reduced GFL power decreases the contribution from the GFL equivalent frequency (third column) due to the reduction in the GFL current injection. For the angle part in (\ref{eq:HeatOprt_Angl}), the coefficients in the third column increase, while those in the fourth column decrease because the reduced GFL current results in a smaller angle difference between the GFL and the measurement location. By combining the updated amplitude and angle parts in (\ref{eq:HeatOprt_Volt}) and (\ref{eq:HeatOprt_Angl}), the contributions from the GFL in the last two rows of the final superposition matrix (\ref{eq:HeatOprt_all}) are adjusted.

\begin{subequations}\label{eq:HeatOprt}
\begin{equation}\label{eq:HeatOprt_Volt}
    \boldsymbol{\Psi}_{Am}^{\prime} = 
    \left[
    \renewcommand{\arraycolsep}{2pt}
    \begin{matrix}
        \cellcolor[rgb]{0.000,0.340,0.305} \textcolor{white}{1.03} & \cellcolor[rgb]{0.000,0.340,0.305} \textcolor{white}{1.04} & \cellcolor[rgb]{0.454,0.770,0.724} 0.50 & \cellcolor[rgb]{0.927,0.957,0.953} 0.02 \\
        \cellcolor[rgb]{0.000,0.340,0.305} \textcolor{white}{1.04} & \cellcolor[rgb]{0.000,0.331,0.296} \textcolor{white}{1.05} & \cellcolor[rgb]{0.454,0.770,0.724} 0.51 & \cellcolor[rgb]{0.927,0.957,0.953} 0.02 \\
        \cellcolor[rgb]{0.000,0.323,0.287} \textcolor{white}{1.07} & \cellcolor[rgb]{0.000,0.315,0.278} \textcolor{white}{1.08} & \cellcolor[rgb]{0.439,0.760,0.714} 0.52 & \cellcolor[rgb]{0.927,0.957,0.953} 0.02 \\
    \end{matrix}
    \right]
\end{equation}
\begin{equation}\label{eq:HeatOprt_Angl}
    \boldsymbol{\Psi}_{Ph}^{\prime} = 
    \left[
    \renewcommand{\arraycolsep}{2pt}
    \begin{matrix}
        \cellcolor[rgb]{0.000,0.315,0.278} \textcolor{white}{1.00} & \cellcolor[rgb]{0.000,0.315,0.278} \textcolor{white}{1.00} & \cellcolor[rgb]{0.064,0.456,0.425} \textcolor{white}{0.89} & \cellcolor[rgb]{0.846,0.938,0.926} 0.46 \\
        \cellcolor[rgb]{0.000,0.340,0.305} \textcolor{white}{0.98} & \cellcolor[rgb]{0.000,0.323,0.287} \textcolor{white}{1.00} & \cellcolor[rgb]{0.023,0.419,0.388} \textcolor{white}{0.92} & \cellcolor[rgb]{0.927,0.957,0.953} 0.40 \\
        \cellcolor[rgb]{0.000,0.323,0.287} \textcolor{white}{0.99} & \cellcolor[rgb]{0.000,0.315,0.278} \textcolor{white}{1.00} & \cellcolor[rgb]{0.112,0.504,0.474} \textcolor{white}{0.85} & \cellcolor[rgb]{0.750,0.908,0.885} 0.53 \\
    \end{matrix}
    \right]
\end{equation}
\begin{equation}\label{eq:HeatOprt_all}
    \boldsymbol{A}^{\prime} = 
    \left[
    \renewcommand{\arraycolsep}{2pt}
    \begin{matrix}
        \cellcolor[rgb]{0.000,0.315,0.278} \textcolor{white}{0.73} & \cellcolor[rgb]{0.604,0.857,0.817} 0.24 & \cellcolor[rgb]{0.902,0.952,0.946} 0.02 & \cellcolor[rgb]{0.927,0.957,0.953} 0.0004 \\
        \cellcolor[rgb]{0.409,0.739,0.695} 0.37 & \cellcolor[rgb]{0.054,0.447,0.416} \textcolor{white}{0.60} & \cellcolor[rgb]{0.893,0.950,0.943} 0.03 & \cellcolor[rgb]{0.927,0.957,0.953} 0.0005 \\
        \cellcolor[rgb]{0.198,0.584,0.553} 0.48 & \cellcolor[rgb]{0.267,0.637,0.603} 0.44 & \cellcolor[rgb]{0.846,0.938,0.926} 0.08 & \cellcolor[rgb]{0.927,0.957,0.953} 0.0016 \\
    \end{matrix}
    \right]
\end{equation}
\end{subequations}

\subsubsection{Extended FD Formula for Branch Frequency} The frequencies of branches $01-04$, $02-07$, and $10-11$, along with the frequency sources from SGs and GFLs, are shown in Fig. \ref{fig:CurrAnaly}. The branch frequency superposition matrix and its various determinants are presented in (\ref{eq:HeatCurr_AmB})–(\ref{eq:HeatCurr_all}).

\begin{figure*}[!t]
    \begin{equation}\label{eq:HeatCurr_Angl}\tag{25c}
    \begin{aligned}
        \boldsymbol{\Psi}_{Ph} = &
        \left[
        \renewcommand{\arraycolsep}{2pt}
        \begin{matrix}
            \cos \! \left( \! \delta_{1}^{G} \! - \! \theta_{01-04}^{B}  \! + \! \zeta_{01-04,1}^{B,G} \! \right) & \cos \! \left( \! \delta_{2}^{G} \! - \! \theta_{01-04}^{B} \! + \! \zeta_{01-04,2}^{B,G} \! \right) & \cos \! \left( \! \theta_{1}^{F,I} \! - \! \theta_{01-04}^{B} \! + \! \zeta_{01-04,1}^{B,F} \! \right) & \sin \! \left( \! \theta_{1}^{F,I} \! - \! \theta_{01-04}^{B} \! + \! \zeta_{01-04,1}^{B,F} \! \right) \\
            \cos \! \left( \! \delta_{1}^{G} \! - \! \theta_{02-07}^{B}  \! + \! \zeta_{02-07,1}^{B,G} \! \right) & \cos \! \left( \! \delta_{2}^{G} \! - \! \theta_{02-07}^{B} \! + \! \zeta_{02-07,2}^{B,G} \! \right) & \cos \! \left( \! \theta_{1}^{F,I} \! - \! \theta_{02-07}^{B} \! + \! \zeta_{02-07,1}^{B,F} \! \right) & \sin \! \left( \! \theta_{1}^{F,I} \! - \! \theta_{02-07}^{B} \! + \! \zeta_{02-07,1}^{B,F} \! \right) \\
            \cos \! \left( \! \delta_{1}^{G} \! - \! \theta_{10-11}^{B}  \! + \! \zeta_{10-11,1}^{B,G} \! \right) & \cos \! \left( \! \delta_{2}^{G} \! - \! \theta_{10-11}^{B} \! + \! \zeta_{10-11,2}^{B,G} \! \right) & \cos \! \left( \! \theta_{1}^{F,I} \! - \! \theta_{10-11}^{B} \! + \! \zeta_{10-11,1}^{B,F} \! \right) & \sin \! \left( \! \theta_{1}^{F,I} \! - \! \theta_{10-11}^{B} \! + \! \zeta_{10-11,1}^{B,F} \! \right) \\
        \end{matrix}
        \right] \\
        = &
        \left[
        \renewcommand{\arraycolsep}{2pt}
        \begin{matrix}
            \cellcolor[rgb]{0.641,0.398,0.091} 0.73 & \cellcolor[rgb]{0.913,0.818,0.582} -0.43 & \cellcolor[rgb]{0.352,0.199,0.019} \textcolor{white}{-0.94} & \cellcolor[rgb]{0.960,0.898,0.735} 0.34 \\
            \cellcolor[rgb]{0.972,0.943,0.865} 0.24 & \cellcolor[rgb]{0.844,0.696,0.398} 0.52 & \cellcolor[rgb]{0.641,0.398,0.091} -0.72 & \cellcolor[rgb]{0.681,0.436,0.118} 0.69 \\
            \cellcolor[rgb]{0.661,0.417,0.104} 0.71 & \cellcolor[rgb]{0.859,0.733,0.449} 0.50 & \cellcolor[rgb]{0.778,0.543,0.215} 0.61 & \cellcolor[rgb]{0.550,0.318,0.040} \textcolor{white}{-0.79} \\
        \end{matrix}
        \right]
    \end{aligned}
    \end{equation}
\end{figure*}

\begin{subequations}\label{eq:HeatCurr}
\begin{equation}\label{eq:HeatCurr_AmB}
\begin{aligned}
    \boldsymbol{\Psi}_{D} = &
    \left[
    \renewcommand{\arraycolsep}{2pt}
    \begin{matrix}
        D_{01-04, 1}^{B,G} & D_{01-04, 2}^{B,G} & D_{01-04, 1}^{B,F} & D_{01-04, 1}^{B,F} \\
        D_{02-07, 1}^{B,G} & D_{02-07, 2}^{B,G} & D_{02-07, 1}^{B,F} & D_{02-07, 1}^{B,F} \\
        D_{10-11, 1}^{B,G} & D_{10-11, 2}^{B,G} & D_{10-11, 1}^{B,F} & D_{10-11, 1}^{B,F} \\
    \end{matrix}
    \right] \\
    = & 
    \left[
    \renewcommand{\arraycolsep}{2pt}
    \begin{matrix}
        \cellcolor[rgb]{0.352,0.199,0.019} \textcolor{white}{2.73} & \cellcolor[rgb]{0.651,0.407,0.097} 1.95 & \cellcolor[rgb]{0.969,0.926,0.812} 0.46 & \cellcolor[rgb]{0.969,0.926,0.812} 0.46 \\
        \cellcolor[rgb]{0.651,0.407,0.097} 1.95 & \cellcolor[rgb]{0.465,0.259,0.017} \textcolor{white}{2.45} & \cellcolor[rgb]{0.971,0.933,0.833} 0.41 & \cellcolor[rgb]{0.971,0.933,0.833} 0.41 \\
        \cellcolor[rgb]{0.972,0.940,0.854} 0.36 & \cellcolor[rgb]{0.972,0.943,0.865} 0.32 & \cellcolor[rgb]{0.930,0.843,0.626} 0.87 & \cellcolor[rgb]{0.930,0.843,0.626} 0.87 \\
    \end{matrix}
    \right]
\end{aligned}
\end{equation}
\begin{equation}\label{eq:HeatCurr_Volt}
\begin{aligned}
    \boldsymbol{\Psi}_{Am} = &
    \left[
    \renewcommand{\arraycolsep}{2pt}
    \begin{matrix}
        \frac{E_{1}^{G0}}{I_{01-04}^{B0}} & \frac{E_{2}^{G0}}{I_{01-04}^{B0}} & \frac{I_{1}^{F0}}{I_{01-04}^{B0}} & \frac{1}{\omega^{0} I_{01-04}^{B0}} \\
        \frac{E_{1}^{G0}}{I_{02-07}^{B0}} & \frac{E_{2}^{G0}}{I_{02-07}^{B0}} & \frac{I_{1}^{F0}}{I_{02-07}^{B0}} & \frac{1}{\omega^{0} I_{02-07}^{B0}} \\
        \frac{E_{1}^{G0}}{I_{10-11}^{B0}} & \frac{E_{2}^{G0}}{I_{10-11}^{B0}} & \frac{I_{1}^{F0}}{I_{10-11}^{B0}} & \frac{1}{\omega^{0} I_{10-11}^{B0}} \\
    \end{matrix}
    \right] \\
    = &
    \left[
    \renewcommand{\arraycolsep}{2pt}
    \begin{matrix}
        \cellcolor[rgb]{0.363,0.204,0.018} \textcolor{white}{1.29} & \cellcolor[rgb]{0.352,0.199,0.019} \textcolor{white}{1.30} & \cellcolor[rgb]{0.560,0.327,0.045} \textcolor{white}{1.02} & \cellcolor[rgb]{0.972,0.940,0.854} 0.02 \\
        \cellcolor[rgb]{0.793,0.570,0.244} 0.67 & \cellcolor[rgb]{0.786,0.556,0.229} 0.67 & \cellcolor[rgb]{0.844,0.696,0.398} 0.53 & \cellcolor[rgb]{0.972,0.943,0.865} 0.01 \\
        \cellcolor[rgb]{0.420,0.233,0.016} \textcolor{white}{1.21} & \cellcolor[rgb]{0.409,0.227,0.017} \textcolor{white}{1.22} & \cellcolor[rgb]{0.601,0.361,0.067} \textcolor{white}{0.96} & \cellcolor[rgb]{0.972,0.940,0.854} 0.02 \\
    \end{matrix}
    \right]
\end{aligned}
\end{equation}
\begin{equation}\label{eq:HeatCurr_all}\tag{25d}
\begin{aligned}
    \boldsymbol{A} = & \boldsymbol{\Psi}_{D} \times \boldsymbol{\Psi}_{Am} \times\boldsymbol{\Psi}_{Ph} \\
    = & \left[
    \renewcommand{\arraycolsep}{1pt}
    \begin{matrix}
        A_{01-04,1}^{B,G} & A_{01-04,2}^{B,G} & A_{01-04,1}^{B,F,Ph} & A_{01-04,1}^{B,F,Am} \\
        A_{02-07,1}^{B,G} & A_{02-07,2}^{B,G} & A_{02-07,1}^{B,F,Ph} & A_{02-07,1}^{B,F,Am} \\
        A_{10-11,1}^{B,G} & A_{10-11,2}^{B,G} & A_{10-11,1}^{B,F,Ph} & A_{10-11,1}^{B,F,Am} \\
    \end{matrix}
    \right] \\
    = &
    \left[
    \renewcommand{\arraycolsep}{2pt}
    \begin{matrix}
        \cellcolor[rgb]{0.352,0.199,0.019} \textcolor{white}{2.55} & \cellcolor[rgb]{0.829,0.655,0.345} -1.11 & \cellcolor[rgb]{0.951,0.880,0.697} -0.44 & \cellcolor[rgb]{0.972,0.943,0.865} 0.003 \\
        \cellcolor[rgb]{0.964,0.907,0.758} 0.30 & \cellcolor[rgb]{0.876,0.764,0.494} 0.85 & \cellcolor[rgb]{0.969,0.926,0.812} -0.16 & \cellcolor[rgb]{0.972,0.943,0.865} 0.003 \\
        \cellcolor[rgb]{0.964,0.907,0.758} 0.30 & \cellcolor[rgb]{0.968,0.923,0.801} 0.19 & \cellcolor[rgb]{0.944,0.866,0.669} 0.50 & \cellcolor[rgb]{0.972,0.943,0.865} -0.013 \\
    \end{matrix}
    \right]
\end{aligned}
\end{equation}
\end{subequations}

For the amplitude of the current superposition matrix in (\ref{eq:HeatCurr_AmB}), the numerical values differ from those in (\ref{eq:CaseHeat_AmB}) due to the multiplication by line admittance in (\ref{eq:CurrSup_DTG_sta})–(\ref{eq:CurrSup_DTF_end}), which makes the forms of $\dot{D}^{B,G}$ and $\dot{D}^{B,F}$ more complex. Similarly, for the angle part, including the angles of $\dot{D}^{B,G}$ and $\dot{D}^{B,F}$ as given in (\ref{eq:HeatCurr_Angl}), the values can be negative, which is a notable distinction from (\ref{eq:CaseHeat_Angl}). As a result, the final superposition matrix in (\ref{eq:HeatCurr_all}) exhibits the following characteristics: (i) it is determined by the electrical distance between the measured branch and the source generator; (ii) The contribution from GFLs is smaller than that from SGs, particularly for the GFL current amplitude $\mathrm{d} I^{F}/\mathrm{d} t$; (iii) the frequency superposition coefficients can be negative, or certain sources may have contribution coefficients exceeding $1$, as previously discussed. As shown in Fig. \ref{fig:CurrAnaly}, the branch frequency exhibits more complex dynamics that are not entirely governed by the SG rotor frequency. Essentially, the branch frequency is not fully bounded by the frequency sources, which is fundamentally different from the node frequency.

\section{Conclusion\label{Sec:Conclu}}

Part I of this series examines the effect of GFLs on system frequency spatial variation. An extended FD formula is first derived. The network node frequency is shown to be superimposed with contributions from the rotor frequencies of SGs and the equivalent frequencies of GFLs, which originate from the angle and amplitude parts of the GFL interfacing state variable. Similarly, an FD formula for branch current frequency is also derived. By applying time derivatives of phasors and algebraic transformations, the proposed FD formula is derived without any simplifications but remain in a linear form. It is found that the impact of GFLs on network frequency is determined by network topology, parameters, and system power flow. Each determinant can be analyzed quantitatively when system conditions are known. Finally, simulation results reveal errors in existing methods for calculating network node frequency. They confirm that the contribution of GFLs is relatively weak compared to SGs and that the superposition matrix is time-varying. Interestingly, the superposition pattern for branch frequency differs significantly, as it is not bounded by the frequencies of the boundary generators.



\bibliographystyle{IEEEtran}

\bibliography{GFL_Frequency_Dynamics_Part_II}

\begin{thebibliography}{10}
\providecommand{\url}[1]{#1}
\csname url@samestyle\endcsname
\providecommand{\newblock}{\relax}
\providecommand{\bibinfo}[2]{#2}
\providecommand{\BIBentrySTDinterwordspacing}{\spaceskip=0pt\relax}
\providecommand{\BIBentryALTinterwordstretchfactor}{4}
\providecommand{\BIBentryALTinterwordspacing}{\spaceskip=\fontdimen2\font plus
\BIBentryALTinterwordstretchfactor\fontdimen3\font minus
  \fontdimen4\font\relax}
\providecommand{\BIBforeignlanguage}[2]{{%
\expandafter\ifx\csname l@#1\endcsname\relax
\typeout{** WARNING: IEEEtran.bst: No hyphenation pattern has been}%
\typeout{** loaded for the language `#1'. Using the pattern for}%
\typeout{** the default language instead.}%
\else
\language=\csname l@#1\endcsname
\fi
#2}}
\providecommand{\BIBdecl}{\relax}
\BIBdecl

\bibitem{BTan22}
B.~Tan, J.~Zhao, M.~Netto, V.~Krishnan, V.~Terzija, and Y.~Zhang, ``Power
  system inertia estimation: Review of methods and the impacts of
  converter-interfaced generations,'' \emph{Int. J. Elect. Power Energy Syst.},
  vol. 134, p. 107362, 2022.

\bibitem{XChen24}
X.~Chen, Y.~Jiang, V.~Terzija, and C.~Lu, ``Review on measurement-based
  frequency dynamics monitoring and analyzing in renewable energy dominated
  power systems,'' \emph{Int. J. Elect. Power Energy Syst.}, vol. 155, p.
  109520, 2024.

\bibitem{YLiu16}
Y.~Liu, L.~Zhan, Y.~Zhang, P.~N. Markham, D.~Zhou, J.~Guo, Y.~Lei, G.~Kou,
  W.~Yao, J.~Chai, and Y.~Liu, ``Wide-area-measurement system development at
  the distribution level: An fnet/grideye example,'' \emph{IEEE Trans. Power
  Deliv.}, vol.~31, no.~2, pp. 721--731, 2016.

\bibitem{MWAltaf22}
M.~W. Altaf, M.~T. Arif, S.~Saha, S.~N. Islam, M.~E. Haque, and A.~M.~T. Oo,
  ``Effective rocof-based islanding detection technique for different types of
  microgrid,'' \emph{IEEE Trans. Ind. Appl.}, vol.~58, no.~2, pp. 1809--1821,
  2022.

\bibitem{MSun21}
M.~Sun, G.~Liu, M.~Popov, V.~Terzija, and S.~Azizi, ``Underfrequency load
  shedding using locally estimated rocof of the center of inertia,'' \emph{IEEE
  Trans. Power Syst.}, vol.~36, no.~5, pp. 4212--4222, 2021.

\bibitem{FMilano18Review}
F.~Milano, F.~Dörfler, G.~Hug, D.~J. Hill, and G.~Verbi\'{c}, ``Foundations
  and challenges of low-inertia systems (invited paper),'' in \emph{2018 Power
  Systems Computation Conference (PSCC)}, 2018, pp. 1--25.

\bibitem{FMilano17}
F.~Milano and {\'A}.~Ortega, ``Frequency divider,'' \emph{IEEE Trans. Power
  Syst.}, vol.~32, no.~2, pp. 1493--1501, 2017.

\bibitem{XHe21}
X.~He, H.~Geng, and G.~Mu, ``Modeling of wind turbine generators for power
  system stability studies: A review,'' \emph{Renew. Sustain. Energy Rev.},
  vol. 143, 2021.

\bibitem{FMilano19}
F.~Milano and {\'A}.~Manjavacas, Ortega, ``Frequency-dependent model for
  transient stability analysis,'' \emph{IEEE Trans. Power Syst.}, vol.~34,
  no.~1, pp. 806--809, 2019.

\bibitem{JZhang24}
J.~Zhang, J.~Wang, N.~Zhang, P.~Wang, Y.~Wang, and C.~Fang, ``Droop coefficient
  placements for grid-side energy storage considering nodal frequency
  constraints under large disturbances,'' \emph{Appl. Energy}, vol. 357, p.
  122444, 2024.

\bibitem{AHussain21}
A.~Hussain, S.~Hasan, S.~Patil, and W.~Shireen, ``Fast frequency regulation in
  islanded microgrid using model-based load estimation,'' \emph{IEEE Trans.
  Energy Convers.}, vol.~36, no.~4, pp. 3188--3198, 2021.

\bibitem{QMa23}
Q.~Ma, L.~Chen, L.~Li, Y.~Min, Y.~Gong, and K.~Liang, ``Effect of
  grid-following vsc on terminal frequency,'' \emph{IEEE Trans. Power Syst.},
  vol.~38, no.~2, pp. 1775--1778, 2023.

\bibitem{QMa24}
Q.~Ma, L.~Chen, L.~Li, Y.~Min, and Y.~Shi, ``Effect of grid‐following vscs on
  frequency distribution of power grid,'' \emph{IET Renew. Power Gener.},
  vol.~18, no.~14, pp. 2619--2628, 2024.

\bibitem{BTan2022FDF}
B.~Tan, J.~Zhao, N.~Duan, D.~A. Maldonado, Y.~Zhang, H.~Zhang, and M.~Anitescu,
  ``Distributed frequency divider for power system bus frequency online
  estimation considering virtual inertia from dfigs,'' \emph{IEEE J. Emerg.
  Sel. Top. Circuits Syst.}, vol.~12, no.~1, pp. 161--171, 2022.

\bibitem{FMilano22}
F.~Milano, ``Complex frequency,'' \emph{IEEE Trans. Power Syst.}, vol.~37,
  no.~2, pp. 1230--1240, 2022.

\bibitem{DMoutevelis24}
D.~Moutevelis, J.~Rold\'{a}n-P\'{e}rez, M.~Prodanovic, and F.~Milano,
  ``Taxonomy of power converter control schemes based on the complex frequency
  concept,'' \emph{IEEE Trans. Power Syst.}, vol.~39, no.~1, pp. 1996--2009,
  2024.

\bibitem{FMilano20pt1}
F.~Milano and {\'A}.~Ortega, ``A method for evaluating frequency regulation in
  an electrical grid – part i: Theory,'' \emph{IEEE Trans. Power Syst.},
  vol.~36, no.~1, pp. 183--193, 2021.

\bibitem{FMilano20pt2}
{\'A}.~Ortega and F.~Milano, ``A method for evaluating frequency regulation in
  an electrical grid – part ii: Applications to non-synchronous devices,''
  \emph{IEEE Trans. Power Syst.}, vol.~36, no.~1, pp. 194--203, 2021.

\bibitem{JNutaro12}
J.~Nutaro and V.~Protopopescu, ``Calculating frequency at loads in simulations
  of electro-mechanical transients,'' \emph{IEEE Trans. Smart Grid}, vol.~3,
  no.~1, pp. 233--240, 2012.

\bibitem{IEEEC37}
\emph{IEEE Standard for Synchrophasor Measurements for Power Systems}, IEEE
  Standard C37.118.1, 2011.

\bibitem{MZhang18}
M.~Zhang, X.~Yuan, and J.~Hu, ``Inertia and primary frequency provisions of
  pll-synchronized vsc hvdc when attached to islanded ac system,'' \emph{IEEE
  Trans. Power Syst.}, vol.~33, no.~4, pp. 4179--4188, 2018.

\bibitem{AddDoc}
\BIBentryALTinterwordspacing
Appendix for ``revisiting the effect of grid-following converter on frequency
  dynamics''. [Online]. Available:
  \url{https://www.researchgate.net/publication/388122216}
\BIBentrySTDinterwordspacing

\bibitem{JLiu23}
J.~Liu, C.~Wang, J.~Zhao, and T.~Bi, ``Rocof constrained unit commitment
  considering spatial difference in frequency dynamics,'' \emph{IEEE Trans.
  Power Syst.}, vol.~39, no.~1, pp. 1111--1125, 2023.

\end{thebibliography}


\end{document}